\documentstyle[12pt]{article}

\newcommand{\tnote}[1]{}

\def\pmb#1{\setbox0=\hbox{#1}%
 \kern-.025em\copy0\kern-\wd0
 \kern.05em\copy0\kern-\wd0
 \kern-.025em\raise.0433em\box0}

\begin{document}

\begin{titlepage}
\null\vspace{-62pt}
\begin{flushright}
Imperial/TP/95-96/28\\
{\tt hep-ph/9603?}\\
{\LaTeX}-ed on \today\\
\end{flushright}
\vspace{0.5in}
\baselineskip=36pt
\begin{center}
{{\LARGE \bf The Densities, Correlations and Length Distributions of Vortices
Produced at a Gaussian Quench}}\\
\baselineskip=18pt
\vspace{1in}
{\large G.\ Karra and R.\ J.\ Rivers}\\
\vspace{.1in}
{{\it Blackett Laboratory, Imperial College, London SW7 2BZ}}\\

\end{center}

\vfill

\begin{abstract}
We present a model for the formation of relativistic global
vortices (strings)
at a quench, and calculate their density and correlations. The significance
of these results to early universe and condensed-matter
physics is discussed.  In particular, there is always open, or infinite,
string.
\end{abstract}

\end{titlepage}

\section{Introduction}

Many systems produce topological defects, in the form of vortices or monopoles,
on undergoing a phase transition to an ordered state.  In this paper we shall
attempt
to calculate how such topological debris is produced in simple transitions.

As might have been anticipated, the distributions that we shall find are
largely generic,
roughly independent both of initial conditions and of the way in which the
transition is implemented.
It is for this reason that, hitherto, it has not been thought necessary to {\it
derive} defect
distributions in any detail.  In fact, given our poor understanding of the
underlying theories,
it has been a consolation to be able to fall back on generic scaling solutions.
For example,
the large-scale structure of
the universe has been attributed \cite{kibble1,shellard}
to cosmic strings (vortices in the
fields) formed at the Grand Unification era.  Although we have only a primitive
understanding of
the relevant field theories,
it has been suggested that the initial conditions of any
string network are largely washed out after a few expansion times,
at which the network is assumed to approach a scaling regime with a
few large loops and long strings per horizon volume continuing to produce
smaller
loops by self and mutual intersection.

Despite this, there are two circumstances in which this lack of
detailed initial information leaves us at a loss, to which this
paper is largely addressed.  The first, of less interest,
concerns the {\it density} of the defects formed.  We have known for
some time how to make
reasonable qualitative estimates\cite{kibble1} of densities and in this paper
we can confirm
them quantitatively.
Secondly, it is not entirely true to say that the details of the
string distributions are unimportant.  In particular, for the astrophysicists
the scaling solutions
for the early universe mentioned above arguably require some open or
{\it 'infinite'} string, {\it i.e.}, vortices that do not self-intersect.  The
presence of such
string is, in part, determined by the initial conditions and it is
important to know whether it is present for reasonable models and if
so, how much.  For
example, it has been suggested that vortices produced by bubble
nucleation in a strong first-order transition will {\it only} form small
loops\cite{borrill}.

The paper is organised as follows.  We begin by reiterating the main tactics
for determining defect densities and distributions and then
display the forms that they will take in a Gaussian approximation for
the underlying field distributions.
The main tools are the correlation functions of the defect densities.
As it stands, some work is necessary to convert them into easily
measurable, or easily identifiable, quantities.  We present
examples, motivated both by our dynamical model (but simpler) and by current
numerical simulations in astrophysics, to help us understand them better.

In our model these correlation functions are given a concrete realisation in
terms of exponentially growing unstable modes when the phase transition is
implemented by an
{\it instantaneous} quench.  Although this is an unrealisable idealisation,
these results are
used as a benchmark for
more general transitions when, later, we vary both the initial conditions and
the way in
which the quench is implemented. It was already implicit in our earlier
work\cite{alray}, of which
this is a continuation, that for a very fast quench the initial conditions
will, in general, only
determine the subleading behaviour of the defect production.  We
shall extend the work presented there to look for exceptions to this general
behaviour.
Further, it will be seen
that changing the {\it rate} at which the
quench is implemented can be approximately equivalent to changing the {\it
time}
at which the transition can be said to have begun.
As a result, the preliminary work of \cite{alray}
for  instantaneous quenches can be extended  with only minor modification.

After a discussion of the way in which defects freeze into the field we
conclude with
some observations
about the length
distribution of vortices, obtained by interpreting some recent
numerical simulations\cite{andy,andy2} in the light of our model.
It will be seen that, whatever we do, there will be infinite string.
Yet again we
assume flat spacetime, for simplicity.  Our conclusions are
thus, in this regard, more applicable to weak coupling condensed matter physics
than
the early universe.  Interestingly, in a condensed matter context, the
same correlation functions should enable us to
estimate the superflow that would occur at a superfluid
quench from fluctuations alone\cite{zurek1}.  We shall make some
steps in this direction.

\section{Defect Distributions}

Before going into details, some generalities about
defect production will be useful.  Our main interest is in vortices
and our discussion will be centred about them.  As will be seen,
other defects are similar but simpler.
The mechanism for vortex formation (termed the Kibble
mechanism\cite{kibble1} in astroparticle physics) is well understood at a
qualitative level.
In this paper we shall only consider the simplest theory that permits vortices,
that
of a complex scalar field $\phi ({\bf x},t)$.  The complex order
parameter of the theory is $\langle\phi\rangle = \eta e^{i\alpha}$ and the
theory possesses
a global $O(2)$ symmetry that we take to be broken
at its phase transition.
Initially, we take the system to be in the symmetry-unbroken
(disordered) phase, in which the field is distributed about $\phi = 0$
with zero mean.
We assume that, at some time $t = t_0$,
the $O(2)$ symmetry of the ground-state (vacuum) is broken by a rapid change in
the
environment inducing an explicit time-dependence in the field
parameters. Once this quench is completed the $\phi$-field potential
takes the familiar symmetry-broken form
$V(\phi ) = -M^2|\phi^2 | + \lambda |\phi^4 |/4$
with $M^{2}> 0$.

The transition for such a global symmetry is {\it continuous} and
 we expect that, as the complex scalar field begins to
fall from the false ground-state into the true ground-state,
different points on the ground-state manifold (the circle $S^{1}$,
labelled by the phase $\alpha$ of $\langle\phi\rangle$) will be chosen at each
point in space
\footnote{Had it been a first-order transition the field would have
tunnelled towards the groundstate, with very different consequences
for densities and correlations than those described here}.
If this is so then continuity and single valuedness will sometimes force the
field to
remain in the false ground-state at $\phi = 0$ . For example, the phase of
the field may change by an integer multiple of $2 \pi$ on going round
a loop in space. This requires at least one
{\it zero} of the field within the loop, each of which has topological
stability
and characterises a vortex (or string).  As to the density of the
strings, if the phase $\alpha$ is correlated over a distance $\xi$,
then the density of strings passing through any surface will be
$O(\xi^{-2})$ {\it i.e.} a fraction of a string per unit correlation area.
The assumed lack of correlation of the field phase over larger
distances  than $\xi$ (inevitable on causal grounds in cosmological
models) is interpreted as saying that the power $P(k)$ of the field
fluctuations, defined by
\begin{equation}
\langle{\tilde \phi}({\bf k}){\tilde \phi}^{*}({\bf k}')\rangle =
(2\pi )^{3}P(k)\delta^{(3)}({\bf
k}-{\bf k}')
\end{equation}
has the form $P(k) = O(k^{0})$ for small $k$.  That is, $P(k)$
describes {\it white noise} at large distances.  Several numerical
simulations based on this assumption have been performed.  The
standard simulation, by Vachasparti and Vilenkin\cite{tanmay},
assumes a regular cubic lattice, in whose cells the $O(2)$ field
phase is chosen at random.  This assumption of 'white noise' leads
to approximately 80\% of the string network being in open string.
While this number is lattice dependent\cite{mark} there is no doubt
that a substantial fraction of string is open, or 'infinite'.
We shall take these white noise field fluctuation predictions as a further
benchmark against which to compare our dynamical predictions.
On the completion of the
transition, when fluctuations are too weak to eliminate or create strings,
a network of strings survives whose further evolution is
determined by classical considerations as the  field gradients
adjust to minimise the energy.

Vortices are not the only defects permitted by global symmetries
although, ultimately, they are the only ones of  real interest to
us.  More generally, a global $O(N)$ scalar theory permits defects
with integer topological charge in $D = N$ spatial dimensions
(monopoles) or $D = N+1$ spatial dimensions (vortices). Because of their
relative complexity it is helpful to
try out our methods on monopoles, which include the {\it kinks} on the line
for a real scalar field theory ($N=D=1$) as a special case, and the {\it
monopoles} in the
plane for a complex scalar $O(2)$ theory ($N=D=2$). [The case of $O(3)$ global
monopoles in three-dimensional space was examined by us elsewhere
\cite{alray} and we shall not consider it further].  With minor
qualifications\footnote{Because of the peculiarities of one spatial dimension,
that would confuse the issue, we pretend that we are examining a
one-dimensional section of a real field in higher dimensions and
ignoring other degrees of freedom.}
a similar mechanism of continuous phase (or field) separation as that for
vortex production is equally valid for the formation of these
defects also.

The question is, how can we infer these vortex, and other defect,
densities and the density correlations from the microscopic field dynamics?
The answer lies in the fact
noted earlier, that for the global $O(2)$ theory of a complex scalar field
$\phi = (\phi_1 + \phi_2)/\sqrt{2}$, ($\phi_1 ,\phi_2$ real) the
string core is a line of zeroes of the fields $\phi_{a}$ ($a$=1,2).
The characterisation of an $O(N)$ global defect  by its field zeroes
is equally true for vortices in the plane ($N=2$) and kinks on the line
($N=1$).

The problem
is solved if we can identify those zeroes which will freeze
out to define the late-time defects.
This will require careful winnowing, since it is apparent that quantum
fluctuations lead to zeroes
of the fields on all distance scales (even in the disordered phase).
For the moment we ignore this difficulty, and attempt to count {\it every}
zero.
The problem then reduces to that of determining the distribution of
field zeroes, given the distribution of fields. This has been
discussed in the literature on several occasions.  We shall call
repeatedly on the work of Halperin \cite{halperin} and Mazenko and
Liu \cite{maz}, but see also Bray \cite{bray}.

\subsection{Kinks on the line}

As a prologue to the more difficult problem of vortices in three dimensions, we
begin with the much simpler problem of identifying the zeroes of a real field
$\phi (x,t)$
in {\it one}
\footnote{See earlier footnote.}
space dimension.
To see how to proceed,
consider an ensemble of systems evolving from one of a set of disordered
states whose relative probabilities are known,
to an ordered state as indicated above.
At any given time $t$, the field will adopt one of the possible configurations
$\Phi (x)$,
whose zeroes we wish to track.  As the field evolves to its equilibrium values
(one of the two minima of its potential
$V(\phi ) = -M^2\phi^2 + \lambda\phi^4/4$)
these zeroes fluctuate and annihilate, but some of them will come to define the
positions $x$ of
'kinks' (field interpolations from one minimum to the other at which $\Phi '(x)
>0$), some the position of 'antikinks'
(at which $\Phi '(x) <0$), the one-dimensional counterparts of
vortices and 'anti'-vortices.

Suppose, at a given time, the zeroes of $\Phi (x)$ occur at $x = x_1
,x_2, ...$.
It is useful to define {\it two} densities.
The first,
\begin{equation}
{\bar\rho}(x) = \sum_{i} \delta (x - x_{i}),
\label{brho}
\end{equation}
is the {\it total} density of zeroes, not distinguishing between
kink zeroes and
antikink zeroes (by which we now mean zeroes at which the field has
positive or negative derivative). The second is the {\it topological} density,
\begin{equation}
\rho (x) = \sum_{i} n_{i}\delta (x - x_{i}),
\label{rho}
\end{equation}
where $n_{i} = {\rm sign}(\Phi '(x_{i}))$, measuring (net) topological
charge, the number of kink minus the number of antikink zeroes.

Equivalently, in terms of the $\Phi$-field, the total density is
\begin{equation}
{\bar\rho}(x) = \delta [\Phi (x)]|\Phi '(x)|,
\label{rhob}
\end{equation}
since $\Phi '(x)$ is the Jacobian of the transformation from zeroes
to fields.  Similarly, the topological density is
\begin{equation}
\rho (x) = \delta [\Phi (x)]\Phi '(x).
\end{equation}

Analytically,
it is not possible to keep track of individual transitions, but we
can construct ensemble averages.
If the phase change begins
at time $t_{0}$ then, for $t > t_{0}$, it is  possible in principle
to calculate the probability $p_{t}[\Phi]$ that
$\phi (x, t)$
takes the value $\Phi (x)$ at time $t$.
Ensemble averaging $\langle F[\Phi ]\rangle_{t}$ at time $t$ is understood
as averaging over the field probabilities
$p_{t}[\Phi ]$.
This is not
thermal averaging since we are out of equilibrium.

The situation we have in mind is one in which, for early times after
the transition when
the available space  permits many domains,
\begin{equation}
\langle\rho (x)\rangle_{t} = 0,
\end{equation}
{\it i.e.}an equal likelihood of a kink zero or an antikink zero occurring in
an infinitesimal length, compatible with an initially
disordered state.  However, the total zero density
\begin{eqnarray}
{\bar n}(t)&=& \langle {\bar \rho} (x)\rangle_{t}
\nonumber
\\
&=&\int {\cal D}\Phi\,\,p_{t}[\Phi]\delta [\Phi (x)]|\Phi '(x)|\,\,>\, 0
\label{n1}
\end{eqnarray}
is positive.
The distribution of the zeroes is given by the density
correlation function
\begin{eqnarray}
C(x ;t) &=& \langle\rho (x)\rho (0)\rangle_{t}
\nonumber
\\
&=&\int {\cal D}\Phi\,\,p_{t}[\Phi]\delta [\Phi (x)]\delta [\Phi (0)] \Phi
'(x)\Phi '(0),
\label{c1}
\end{eqnarray}
($x\neq 0$) which will also be non-zero.

To make this relationship
more concrete, we observe that, on separating out the diagonal and non-diagonal
terms
in the expansion for $\rho (x)\rho (y)$ from (\ref{rho}), then
\begin{equation}
\rho (x)\rho (y) = {\bar \rho}(x)\delta (x-y) + g(x-y)
\end{equation}
where
\begin{equation}
g(x-y) = \sum_{i\neq j}\,n_{i}n_{j}\delta (x-x_{i})\delta (y-x_{j}).
\end{equation}
That is,
\begin{equation}
\langle\rho (x)\rho (0)\rangle_{t} = {\bar n}(t)\delta (x) + C(x;t).
\end{equation}
where $C(x;t) = \langle g(x)\rangle_{t}$.  Charge conservation
\begin{equation}
\int_{-\infty}^{\infty}dx\,\langle\rho (x)\rho (0)\rangle_{t} = 0
\end{equation}
implies
\begin{equation}
\int_{-\infty}^{\infty}dx\,C(x;t) = -{\bar n}(t),
\label{Cint}
\end{equation}
requiring that $C(x;t)=C(-x;t)$ be largely negative.

We can now relate  $C(x;t)$ to the distribution and spacing of
zeroes by calculating the variance of the topological charge
\begin{equation}
n_{L} = \int_{0}^{L}dx\,\rho (x)
\end{equation}
on the interval $I = [0,L]$.  In this particularly simple case
$n_{L} = -1, 0, 1$. Then
\begin{eqnarray}
(\Delta_{t} n_{L})^{2}&=&\langle n_{L}^{2}\rangle_{t}
\nonumber
\\
&=& \int_{0}^{L}dx\int_{0}^{L}dy\,\langle\rho (x)\rho (y)\rangle_{t}
\nonumber
\\
&=& L{\bar n}(t) + \int_{0}^{L}dx\int_{0}^{L}dy\,C(x-y;t)
\nonumber
\\
&=&- \int_{x<0}^{x>L}dx\int_{0}^{L}dy\,C(x-y;t).
\label{var2}
\end{eqnarray}
from (\ref{Cint}).

If $h(x;t)$ is defined by $C(x;t) = \partial
h(x;t)/\partial x$ then
\begin{equation}
\frac{\partial (\Delta_{t} n_{L})^{2}}{\partial L} = -2h(L;t).
\end{equation}
However, if $p_{L}$ is the probability that, on average, a length
$L$ of the line contains an odd number of zeroes, then $(\Delta_{t}
n_{L})^{2}=p_{L}$.
Thus
\begin{equation}
C(L;t)=h'(L;t)=-\frac{1}{2}\frac{d^{2}p_{L}}{dL^{2}}.
\end{equation}
As an extreme case we note that, for
an array of equally spaced zeroes, separation $\xi (t)$, $p_{L}$ is
saw-toothed, period $2\xi$, from which
\begin{equation}
h(x;t) = \frac{1}{2\pi\xi (t)}{\rm sign}(\sin(\pi x/\xi (t))),
\label{hr}
\end{equation}
whence $C(x;t) = h'(x;t)$ is a sum of $\delta$-functions.
At the other extreme, independent (Poisson) zeroes, mean separation $\xi (t)$,
give
\begin{equation}
h(x;t) = e^{-2|x|/\xi (t)}/2\xi (t)
\label{poisson}
\end{equation}
and
\begin{equation}
C(L;t)=-e^{-2L/\xi (t)}/\xi^{2}(t).
\end{equation}

These provide a useful guide, and we shall return to them later.
However, some caution is necessary since, on the line, kink zero
is followed by anti-kink zero and vice-versa.
As an intermediate
step between this and the vortices of the complex field $\phi ({\bf x}, t)$ in
three
dimensions, we shall consider the same field in two dimensions, for
which monopoles (point defects) occur.

\subsection{Monopoles in two dimensions}

An $O(2)$ complex field $\Phi ({\bf x})$ in the plane permits
monopoles, at this stage identified by zeroes of  the field with
non-trivial winding number.  After the transition is completed, the relevant
zeroes will define the centres of effectively classical monopoles. Although
Derrick's theorem prohibits finite-energy static monopole solutions to the
classical field equations, a non-zero monopole density provides a cut-off to
the logarithmic tails of the individual monopoles, and there is no problem on
this score.

For the moment we continue to count every field zero, at whatever scale and
however transient. If $\Phi ({\bf x})$ has zeroes at ${\bf
x_1}, {\bf x_2}, {\bf x_3},...$ the total and topological densities
${\bar\rho}({\bf x})$ and $\rho ({\bf x})$ are the straightforward
generalisations of (\ref{brho}) and (\ref{rho}),
\begin{equation}
{\bar\rho}({\bf x}) = \sum_{i} \delta ({\bf x} -{\bf x}_{i}),
\label{brho2}
\end{equation}
and
\begin{equation}
\rho ({\bf x}) = \sum_{i} n_{i}\delta ({\bf x} -{\bf x}_{i}),
\label{rho2}
\end{equation}
where $n_{i} = \pm 1$ is the winding number of the zero (higher
winding numbers are taken as multiple zeroes).

The relationship between zeroes and fields is through the Jacobian, giving
\begin{equation}
\rho ({\bf x}) = \delta^{2}[\Phi ({\bf x})]\epsilon_{jk}\partial_{j}
\Phi_{1}({\bf x}) \partial_{k}\Phi_{2}({\bf x}),\,\,\,i,j=1,2
\label{rhof2}
\end{equation}
where $\epsilon_{12}=-\epsilon_{21}=1$, otherwise zero.
As before we assume that it is  possible in principle
to calculate the probability $p_{t}[\Phi]$ that
$\phi ({\bf x}, t)$ takes the value $\Phi ({\bf x})$ at time $t$
\footnote{Throughout, it will be convenient to decompose $\Phi$ into
real and imaginary parts as $\Phi  = \frac{1}{\sqrt{2}}(\Phi_{1} +
i\Phi_{2})$.  This is because we wish to
track the field as it falls from the unstable ground-state hump at the centre
of the potential to the ground-state manifold in
Cartesian field space.},
from which we define the ensemble averages
$\langle F[\Phi]\rangle_{t}$.
The topological charge average $\langle\rho ({\bf x})\rangle_{t}$ is taken to
be
zero as before.  The non-zero total density and the topological
density correlation functions are defined as for the one-dimensional
case,
\begin{eqnarray}
{\bar n}(t)&=& \langle {\bar \rho} ({\bf x})\rangle_{t}
\nonumber
\\
&=&\int {\cal D}\Phi\,\,p_{t}[\Phi]
\delta^{2}[\Phi ({\bf x})]|\epsilon_{jk}\partial_{j}
\Phi_{1}({\bf x}) \partial_{k}\Phi_{2}({\bf x})|\,\,>\, 0
\label{n2}
\end{eqnarray}
and
\begin{eqnarray}
C(r;t) &=& \langle\rho ({\bf x})\rho ({\bf 0})\rangle_{t}
\nonumber
\\
&=&\int {\cal D}\Phi\,\,p_{t}[\Phi]\delta^{2}[\Phi ({\bf x})]\delta^{2}[\Phi
({\bf 0})]
\epsilon_{jk}\partial_{j}
\Phi_{1}({\bf x}) \partial_{k}\Phi_{2}({\bf x})
\epsilon_{lm}\partial_{l}
\Phi_{1}({\bf 0}) \partial_{m}\Phi_{2}({\bf 0})
\label{c2}
\end{eqnarray}
($r=|{\bf x}|\neq \bf 0$).
Charge conservation again applies, as
\begin{equation}
\int d^{2}x\,C(r;t) = -{\bar n}(t),
\label{Cint2}
\end{equation}
but it is not so easy to give a direct interpretation to $C({\bf
x};t)$ in terms of monopole separations.

Nonetheless, it is still useful to consider the variance in the topological
charge
\begin{equation}
n_{S} = \int_{{\bf x}\in S}d^{2}x\,\rho ({\bf x}),
\end{equation}
this time through a  region $S$ in the plane, (area $s$) since it can be an
observable quantity.
For example, on quenching $^{4}He$ in an annulus $S$, $(\Delta_{t} n_{S})^{2}$
is a measure of the supercurrent generated by the quench
\cite{zurek1}.
Nothing that we have said so far really requires a relativistic
theory, so let us pursue it a little further.
Then,
from (\ref{Cint2}) it
follows that
\begin{eqnarray}
(\Delta_{t} n_{S})^{2}&=&\int_{{\bf x}\in S}d^{2}x\int_{{\bf y}\in
S}d^{2}y\,\langle\rho
({\bf x})\rho ({\bf y})\rangle_{t}
\nonumber
\\
&=&- \int_{{\bf x}\not\in S}d^{2}x\int_{{\bf y}\in
S}d^{2}y\,\,C(|{\bf x}-{\bf y}|;t)
\label{per}
\end{eqnarray}
the two-dimensional counterpart to (\ref{var2}).
Suppose that there are short-range vortex-zero antivortex-zero correlations in
which
 $C(r;t)$ is short-range in $r$, with lengthscale $\xi (t)$.  If $\xi
(t)$ is the only lengthscale then $C(r;t)$ is $O(\xi^{-4}(t))$.

With ${\bf x}$ outside $S$, and ${\bf y}$ inside $S$, all the
contribution to $(\Delta_{t} n_{S})^{2}$ comes from the vicinity of
the boundary of $S$, rather than the whole area.
More precisely, suppose that $S$ is a disc of radius
$L$ \footnote{The extension to an annulus is straightforward.}.
Then (\ref{per}) can be
written as
\begin{equation}
(\Delta_{t} n_{S})^{2} =-2\pi
\int_{0}^{L}r\,dr\int_{L}^{\infty}r'\,dr'\int_{0}^{2\pi}d\theta
\,\,C(R;t)
\label{per2}
\end{equation}
where
\begin{equation}
R^{2} = r^{2}+r'^{2}-2rr'\cos \theta .
\end{equation}
The integration region in $r,r'$ is restricted to $|r'-L| = O(\xi (t)) =
|L-r|$.
If $C(r;t)$ is non-singular at $r=0$ and not varying too rapidly at
the origin, then
the $\theta$-integration is  limited to a range $O(\xi (t)/L)$, whence
\begin{equation}
(\Delta_{t} n_{S})^{2} = O \bigg(\frac{L}{\xi (t)}\bigg).
\end{equation}
This robust result is compatible with, but does not imply, a random
walk in field phase around the perimeter and we shall return to it
later, when we shall see that our model implies short-range
correlations
\footnote{At the other extreme we note that, if the monopoles and
antimonopoles are individually and mutually uncorrelated (a double Poisson
distribution), then
$(\Delta n_{S})^{2} =  s{\bar n}$.
That is, the variance grows linearly with area.}.

\subsection{Vortices in Three Dimensions}

Finally reaching our goal of vortices in three dimensions, we  define the {\it
topological line density} of zeroes $\pmb{$\rho$}(\bf r)$
\cite{halperin,maz} by
\begin{equation}
\pmb{$\rho$}({\bf x}) = \sum_{n}\int ds \frac{d{\bf R}_{n}}{ds}
\delta^{3} [{\bf x} - {\bf R}_{n}(s)].
\label{corrr}
\end{equation}
In (2.1) $ds$ is the incremental length along the line of zeroes ${\bf
R}_{n}(s)$ ($n$=1,2,.. .) and $\frac{d{\bf R}_{n}}{ds}$ is a unit
vector pointing in the direction which corresponds to positive
winding number.  As in the previous case, we begin by counting all
zeroes, anticipating that the relevant ones will become the cores
of the strings of the resulting network. Yet again, a finite string
density once the transition is complete provides the cutoff necessary to
eliminate the infrared
divergent tails of the energy densities that we would expect from
Derrick's theorem.

It follows that, in terms of the zeroes of $\Phi ({\bf x})$,
$\rho_{i}({\bf x})$ can be written as
\begin{equation}
\rho_{i}({\bf x}) = \delta^{2}[\Phi ({\bf x})]\epsilon_{ijk}\partial_{j}
\Phi_{1}({\bf x}) \partial_{k}\Phi_{2}({\bf x}),
\label{rho3}
\end{equation}
where
$\delta^{2}[\Phi ({\bf x})] = \delta[\Phi_{1} ({\bf x})] \delta[\Phi_{2}
({\bf x})]$.
The coefficient of the $\delta$-function in (\ref{rho}) is
the Jacobian of the more complicated transformation from line zeroes to field
zeroes.
We shall also need the {\it total line density} $\bar{\rho}({\bf
x})$, the counterpart of $\bar{\rho}(x)$ of (\ref{rhob}),
\begin{equation}
\bar{\rho_{i}}({\bf x}) = \delta^{2}[\Phi ({\bf x})]|\epsilon_{ijk}\partial_{j}
\Phi_{1}({\bf x}) \partial_{k}\Phi_{2}({\bf x})|.
\label{rhobar}
\end{equation}

As before, ensemble averaging $\langle F[\Phi ]\rangle_{t}$ at time $t$
means averaging over the field probabilities $p_{t}[\Phi ]$
{}.
Again we assume
\begin{equation}
\langle\rho_{j}({\bf x})\rangle_{t} = 0.
\end{equation}
{\it i.e.} an equal likelihood of a string line-zero or an
antistring line-zero passing
through an infinitesimal area.
However,
\begin{equation}
{\bar n}(t) = \; \langle\bar{\rho_{i}}({\bf x})\rangle_{t} \; > 0
\label{n3}
\end{equation}
and measures the {\it total} line-zero density in the direction $i$, without
regard to string orientation.  The isotropy of the initial state
guarantees that ${\bar n}(t)$ is independent of the direction $i$.
Further, the line density
correlation functions
\begin{equation}
C_{ij}({\bf x} ;t) = \langle\rho_{i}({\bf x})\rho_{j}({\bf 0})\rangle_{t}
\end{equation}
will be non-zero, and give information on the persistence length of
line zeroes.
It will be convenient, for later work, to decompose $C_{ij}({\bf x} ;t)$
as ($r = |{\bf x}|$),
\begin{equation}
C_{ij}({\bf x} ;t) = A(r;t)\bigg(\delta_{ij} -\frac{x_{i}x_{j}}{r^{2}}
\bigg)+
B(r;t)\biggl(\frac{x_{i}x_{j}}{r^{2}}\biggr),
\label{ddc}
\end{equation}
The realisations of ${\bar n}(t)$ and $C_{ij}({\bf x} ;t)$ in
terms of $p_{t}[\Phi ]$ are simple generalisations of (\ref{n2}) and
(\ref{c2}) and will not be given explicitly.

Charge conservation now means
\begin{equation}
\int d^{3}x\,\,C_{ij}({\bf x} ;t)=0
\label{Cint3a}
\end{equation}
without any inhomogeneous term.
On taking the trace in (\ref{Cint3a}) it follows that
\begin{equation}
\int d^{3}x\,\,\bigg(2A(r;t) + B(r;t)\bigg) =0
\label{Cint3b}
\end{equation}
We note that, from (\ref{corrr}), the integral of $\rho_{j}({\bf x})$ over an
open
surface $S$ with normal in the $j$-direction
does not measure the winding number along the boundary
of $S$ since each line zero is weighted by the cosine of its angle of
incidence on $S$.  Thus the variance of the winding number
(topological charge) through $S$ is essentially the two-dimensional
problem discussed previously.

There are no simple quantities that can be calculated for line zeroes
(e.g.Poisson strings).  However, we note that, if ${\bf x} =
(0,0,r)$, then
\begin{equation}
C_{11}({\bf x} ;t) = C_{22}({\bf x} ;t) = A(r;t)
\end{equation}
measures the likelihood of there being a string zero (or antistring zero) in
the same direction at separation $r$.  A negative value of $A$ of $O({\bar
n}^{2})$
indicates the presence of vortex-antivortex line zero pairs at separation $r$.
Further, for the same
$r$,
\begin{equation}
C_{33}({\bf x} ;t) = B(r;t)
\end{equation}
measures the tendency of a vortex to bend on a distance $r$
(provided $r$ is sufficiently small for it to be measuring the same
string). Once $B{\bar n}^{2}$ is negligable the line
has bent away from its initial direction.

\section{Ensemble Averaging}

In relating distributions of defects to field fluctuations we have
seen that we need the field probability $p_{t}[\Phi ]$ at all times.
There is no difficulty in writing a formal expression for $p_{t}[\Phi
]$, although its calculation is another matter.  Details are given in
\cite{alray}, and
we shall only provide the briefest recapitulation.

Take $t=t_0$ as our starting time. Suppose that, at $t_0$, the
system is in a pure state, in which the measurement of $\phi$ would
give $\Phi_0({\bf x})$. That is:-
\begin{equation}
\hat{\phi}(t_0,{\bf x}) | \Phi_0,t_0 \rangle = \Phi_0 | \Phi_0,t_0 \rangle.
\end{equation}
The probability $p_{t_f}[\Phi_f]$ that, at time $t_f>t_0$, the
measurement of $\phi$ will give the value $\Phi_f$ is $p_{t_f}[\Phi_f]
= |\Psi_{f0}|^2$, where $\Psi_{f0}$ is the state-functional with the
specified initial condition.  As a path integral
\begin{equation}
\Psi_{f0}  = \int_{\phi(t_0) = \Phi_0}^{\phi(t_f) = \Phi_f} {\cal D} \phi
\, \exp \biggl \{ i S_t [\phi] \biggr \},
\end{equation}
where $S_t [\phi]$ is the (time-dependent) action that desribes how
the field $\phi$ is driven by the environment,
${\cal D} \phi = \prod_{a=1}^N {\cal D} \phi_a$ and spatial
labels have been suppressed.
Specifically, for $t > t_{0}$ the action for the field is taken to be
\begin{equation}
S_t [\phi] = \int d^{D+1}x \biggl (
\frac{1}{2} \partial_{\mu} \phi_a \partial^{\mu} \phi_a - \frac{1}{2}
m^{2}(t) \phi_a^2 - \frac{1}{4} \lambda (t) (\phi_a^2)^2
\biggr ).
\label{St}
\end{equation}
where $m(t)$, $\lambda (t)$ describe the evolution of the parameters
of the theory under external influences, to which the field responds.
The spatial dimension $D=3$ for the relevant case of vortices, and $D=2$
for monopoles in the plane \footnote{$D=1$ for kinks on the line}.
As with $\Phi$, it is convenient to decompose the complex field $\phi$ in terms
of two massive
real scalar fields $\phi_a$, $a=1, 2$ as $\phi = (\phi_1 + i\phi_2)/\sqrt{2}$,
in terms of which $S[\phi ]$ shows a global $O(2)$ invariance, broken by the
mass
term if $m^{2}(t)$ is negative.

It follows that $p_{t_f}[\Phi_f]$ can be
written in the closed time-path form
\begin{equation}
p_{t_f}[\Phi_f] = \int_{\phi_{\pm}(t_0) = \Phi_0}^{\phi_{\pm}(t_f) =
\Phi_f} {\cal D} \phi_+  {\cal D} \phi_-
\, \exp \biggl \{ i \biggl ( S_t [\phi_+]-S_t [\phi_-] \biggr ) \biggr \}.
\end{equation}
Instead of separately integrating $\phi_{\pm}$ along the time paths
$t_0 \leq t \leq t_f$, the integral can be interpreted as
time-ordering of a field $\phi$ along the closed path $C_+ \oplus C_-$
where $\phi =\phi_+$ on $C_+$ and $\phi= \phi_-$ on $C_-$.
The two-field notation is misleading in that it suggests that the
$\phi_{+}$ and $\phi_{-}$ fields are decoupled.  That this is not so
follows immediately from the fact that $\phi_{+}(t_f) =\phi_{-}(t_f)$.
It is necessary to keep this in mind when we extend the contour from $t_f$ to
$t= \infty$.
Either $\phi_+$ or $\phi_-$ is an equally good candidate for the
physical field, but we choose $\phi_+$.
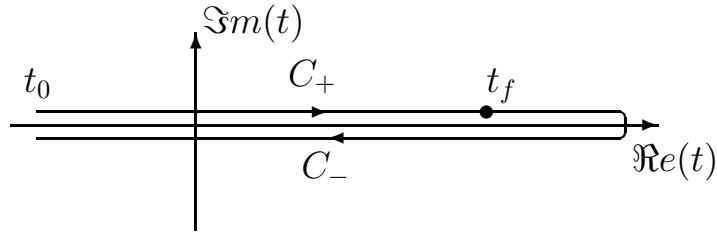
\begin{figure}[htb]
\begin{center}
\setlength{\unitlength}{0.5pt}
\begin{picture}(495,200)(35,600)
\put(260,640){\makebox(0,0)[lb]{\large $C_-$}}
\put(185,750){\makebox(0,0)[lb]{\large $\Im m(t)$}}
\put(250,710){\makebox(0,0)[lb]{\large $C_+$}}
\put(510,645){\makebox(0,0)[lb]{\large $\Re e (t)$}}
\put(400,705){\makebox(0,0)[lb]{\large $t_f$}}
\put( 50,705){\makebox(0,0)[lb]{\large $t_0$}}
\thicklines
\put(400,690){\circle*{10}}
\put( 40,680){\vector( 1, 0){490}}
\put( 60,690){\vector( 1, 0){220}}
\put(280,690){\line( 1, 0){220}}
\put(500,680){\oval(10,20)[r]}
\put(500,670){\vector(-1, 0){220}}
\put(280,670){\line(-1, 0){220}}
\put(180,600){\vector( 0, 1){150}}
\end{picture}
\end{center}
\caption{The closed timepath contour $C_+ \oplus C_-$.}
\end{figure}
With this choice and suitable normalisation, $p_{t_f}$ becomes
\begin{equation}
p_{t_f}[\Phi_f] = \int_{\phi_{\pm}(t_0) = \Phi_0}
{\cal D} \phi_+  {\cal D} \phi_- \, \delta [ \phi_+(t) - \Phi_f ]
\, \exp \biggl \{ i \biggl ( S_t [\phi_+]-S_t [\phi_-] \biggr ) \biggr \},
\end{equation}
where $\delta [ \phi_+(t) - \Phi_f ]$ is a delta functional, imposing
the constraint $\phi_+(t,{\bf x}) = \Phi_f ({\bf x})$ for each ${\bf x}$.

The choice of a pure state at time $t_0$ is too simple to be of any
use. The one fixed condition is that we
begin in a symmetric state with $\langle \phi \rangle = 0$ at time
$t=t_0$. Otherwise, our ignorance is parametrised in the probability
distribution that at time $t_0$, $\phi(t_0,{\bf x}) = \Phi({\bf x})$.
If we allow for an initial probability
distribution $p_{t_0}[\Phi]$ then $p_{t_f}[\Phi_f]$ is generalised to
\begin{equation}
p_{t_f}[\Phi_f] = \int {\cal D} \Phi p_{t_0}[\Phi]
\int_{\phi_{\pm} (t_0) = \Phi} {\cal D} \phi_+  {\cal D}
\phi_- \, \delta [ \phi_+(t_f) - \Phi_f ] \, \exp \biggl \{ i \biggl (
S_t [\phi_+] - S_t [\phi_-] \biggr ) \biggr \}.
\end{equation}
It is impossible to derive $p_t$ analytically for general initial
conditions.  Fortunately, we shall see that, in many circumstances,
the details of the initial condition are largely irrelevant. All the cases
that we might wish to consider are encompassed in the assumption that
$\Phi$ is Boltzmann distributed at time $t_0$ at an effective
temperature of $T_0 = \beta_0^{-1}$ according to a  Hamiltonian
$H_0[\Phi]$, where the subscript {\it zero} denotes $t_0$ rather
than a free field. That is
\begin{equation}
p_{t_0}[\Phi] = \langle \Phi,t_0 | e^{- \beta_0 H_0} | \Phi,t_0 \rangle
= \int_{\phi_3(t_0) = \Phi = \phi_3(t_0-i \beta_0)} {\cal D} \phi_3
\exp \biggl \{ i S_0 [\phi_3] \biggr \},
\end{equation}
for a corresponding action $S_0[\phi_3]$, in which $\phi_3$ is taken
to be periodic in imaginary time with period $\beta_0$. We take
$S_0[\phi_3]$ to have the standard form in $\phi_3$ as
\begin{equation}
S_0[\phi_3] = \int d^{D+1}x \biggl [
\frac{1}{2} (\partial_{\mu} \phi_{3 \, a} )(\partial^{\mu} \phi_{3 \,
a} ) - \frac{1}{2} m_0^2 \phi_{3 \, a}^2 -\frac{1}{4}\lambda_{0}(\phi_{3 \,
a}^2)^{2}
\biggr ].
\label{S-0}
\end{equation}
We stress that $m_0$, $\lambda_{0}$ and $\beta_0$ parametrise our uncertainty
in the
initial conditions. The choice $\beta_0 \rightarrow \infty$ corresponds
to choosing the $p_t[\Phi]$ to be determined by the ground state
functional of $H_0$, for example. For the sake of argument we take
$T_{0} = \beta_{0}^{-1}$ to be a temperature higher than the
transition temperature $T_{c}$ and $m_0 = m(T_0)$ ($m_{0}^{2}>0$) to be the
effective mass at this at this temperature.  Whatever, the effect is to give an
action $S_{3}[\phi]$ in which we are in thermal equilibrium for $t<t_0$ during
which period the mass $m(t)$ and coupling constant $\lambda (t)$ take
the constant values $m_0$ and $\lambda_{0}$ respectively.

We now have the explicit form for $p_{t_f}[\Phi_f]$,
\begin{equation}
p_{t_f}[\Phi_f]
= \int_B {\cal D} \phi_3 {\cal D} \phi_+ {\cal D} \phi_- \, \exp \biggl \{
i S_0[\phi_3] + i ( S[\phi_+] - S[\phi_-] )
\biggr \} \,
\delta [ \phi_+(t_f) - \Phi_f ],
\end{equation}
where the boundary condition $B$ is $\phi_{\pm}(t_0) = \phi_3(t_0) =
\phi_3(t_0- i \beta_0)$. This can be written as the time ordering of a
single field:-
\begin{equation}
p_{t_f} [ \Phi_f] = \int_B {\cal D} \phi \, e^{i S_C [\phi]} \, \delta [
\phi_+ (t_f) - \Phi_f ],
\end{equation}
along the contour $C=C_+ \oplus C_- \oplus C_3$, extended to include a
third imaginary leg, where $\phi$ takes the values $\phi_+$, $\phi_-$
and $\phi_3$ on $C_+$, $C_-$ and $C_3$ respectively, for which $S_C$
is $S[\phi_+]$, $S[\phi_-]$ and $S_0[\phi_3]$.
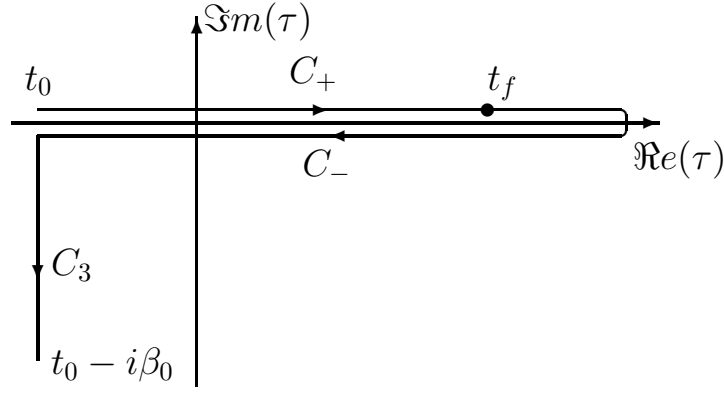
\begin{figure}[htb]
\begin{center}
\setlength{\unitlength}{0.5pt}
\begin{picture}(495,280)(35,480)
\put( 70,565){\makebox(0,0)[lb]{\large $C_3$}}
\put(260,640){\makebox(0,0)[lb]{\large $C_-$}}
\put(185,750){\makebox(0,0)[lb]{\large $\Im m(\tau)$}}
\put(250,710){\makebox(0,0)[lb]{\large $C_+$}}
\put(510,645){\makebox(0,0)[lb]{\large $\Re e (\tau)$}}
\put(400,705){\makebox(0,0)[lb]{\large $t_f$}}
\put( 50,705){\makebox(0,0)[lb]{\large $t_0$}}
\put(70,490){\makebox(0,0)[lb]{\large $t_0-i \beta_0$}}
\thicklines
\put(400,690){\circle*{10}}
\put( 40,680){\vector( 1, 0){490}}
\put( 60,690){\vector( 1, 0){220}}
\put(280,690){\line( 1, 0){220}}
\put(500,680){\oval(10,20)[r]}
\put(500,670){\vector(-1, 0){220}}
\put(280,670){\line(-1, 0){220}}
\put( 60,670){\vector( 0,-1){110}}
\put( 60,560){\line( 0,-1){ 60}}
\put(180,480){\vector( 0, 1){280}}
\end{picture}
\end{center}
\caption{A third imaginary leg}
\end{figure}
Henceforth we drop the suffix $f$ on $\Phi_f$ and take the
origin in time from which the evolution begins as $t_0 =0$.

We perform one final manoeuvre with $p_t[\Phi]$ before resorting to
further approximation to demonstrate how we can average without
having to calculate $p_t[\Phi]$ explicitly. Further,
this will enable us to avoid a nominally ill-defined
inversion of a two-point function later on, a consequence of the
seeming independence of $\phi_{+}$ and $\phi_{-}$ mentioned earlier.
Consider the generating
functional:-
\begin{equation}
Z[j_+,j_-,j_3] = \int_B {\cal D} \phi \, \exp \biggl \{ i S_C[\phi] +
i \int j \phi \biggr \},
\end{equation}
where $\int j \phi$ is a short notation for:-
\begin{equation}
\int j \phi \equiv \int_0^{\infty} dt \, \,  [ \, j_+(t) \phi_+(t) - j_-
\phi_-(t) \, ] \, + \int_0^{-i \beta} j_3(t) \phi_3(t) \, dt,
\end{equation}
omitting spatial arguments. Then introducing $\alpha_a({\bf x})$ where
$a=1, \ldots , N$, we find:-
\begin{eqnarray}
\nonumber
p_{t_f} [\Phi] &=& \int {\cal D} \alpha \int_B {\cal D } \phi \,\,
\exp \biggl \{i
S_C[\phi] \biggr \} \,\,
\exp \biggl \{ i\int d^{D} x \alpha_a({\bf x}) [ \phi_+(t_f,{\bf
x}) - \Phi({\bf x}) ]_a \biggr \}
\\
\nonumber
&=& \int {\cal D} \alpha \, \, \exp \biggl \{ -i \int \alpha_a \Phi_a
\biggr \} \,Z[\overline{\alpha},0,0],
\end{eqnarray}
where $\overline{\alpha}$ is the source $\overline{\alpha}(t,{\bf x}) =
\alpha({\bf x}) \delta (t-t_f)$. As with ${\cal D} \phi$, ${\cal D}
\alpha$ denotes $\prod_1^N {\cal D} \alpha_a$.
Ensemble averages are now expressible in terms of $Z_\mu$.  Of
particular relevance,
$W_{ab}(|{\bf x} -{\bf x} '|;t) = \langle\Phi_{a}({\bf x})\Phi_{b}({\bf
x}')\rangle_{t}$ is given by
\begin{eqnarray}
W_{ab}(|{\bf x} -{\bf x} '|;t) &=& \int {\cal D}\Phi\,\, \Phi_{a}({\bf
x})\Phi_{b}({\bf
x}')
\int {\cal D} \alpha \,\,\exp \biggl \{ -i \int \alpha_a \Phi_a
\biggr \} \,\,Z_{\mu}[\overline{\alpha},0,0]
\nonumber
\\
&=&
-\int {\cal D}\alpha\,\, \frac{\delta^{2}}{\delta\alpha_{a}({\bf
x})\delta\alpha_{b}({\bf
x}')}
\int {\cal D} \Phi\,\,\exp \biggl \{ -i \int \alpha_a \Phi_a
\biggr \}\,\, Z_{\mu}[\overline{\alpha},0,0]
\nonumber
\\
&=&
-\int {\cal D}\alpha\,\,\frac{\delta^{2}}{\delta\alpha_{a}({\bf
x})\delta\alpha_{b}({\bf
x}')}
\biggl \{\delta^{2}\,[ \alpha ]\,\,Z_{\mu}[\overline{\alpha},0,0]\biggr \}
\end{eqnarray}
On integrating by parts
\begin{eqnarray}
W_{ab}(|{\bf x} -{\bf x} '|;t) &=&
-\frac{\delta^{2}}{\delta\alpha_{a}({\bf x})\delta\alpha_{b}({\bf
x}')}Z_{\mu}[\overline{\alpha},0,0]\bigg |_{\alpha = 0}
\nonumber
\\
&=& \langle\phi_{a}({\bf x},t)\phi_{b}({\bf
x}',t)\rangle,
\end{eqnarray}
the equal-time thermal Wightman function with the given thermal boundary
conditions.
Because of the non-equilibrium time evolution
there is no time translation invariance in the double time label.

\section{A Gaussian Model for Defect Distributions}

We have yet to specify the nature of the quench but it is already
apparent that, if we are to make further progress, additional approximations
are necessary.
We return to our one-dimensional example.

\subsection{Kinks}

Suppose the fields and their derivatives at the same point show
approximate independence.  Then the zero density ${\bar n}(t)$ of (\ref{n1})
separates as
\begin{equation}
{\bar n}(t)\approx\langle\delta [\Phi (x)]\rangle_{t} \langle |\Phi '(x)|
\rangle_{t}.
\label{n11}
\end{equation}
and we can estimate, or bound, each factor separately.
Specifically, on writing the $\delta$-function term as
\begin{equation}
\langle\delta [\Phi (x)]\rangle_{t} =
\int d\!\!\!/\alpha\langle \exp\{i\int dy\,j(y)\Phi (y)\}\rangle_{t}
\end{equation}
where $j(y) =\alpha\delta (y-x)$, it follows that
\begin{equation}
{\bar n}(t)\leq
\frac{1}{\sqrt{2\pi}}\biggl |\frac{W''(0;t)}{W(0;t)}\biggr
|^{\frac{1}{2}}\bigg[ 1 + \mbox{Connected\, Correlation\, Functions}\bigg]
\label{napp}
\end{equation}
on using the Schwarz inequality,
where primes on $W$ denote differentiation with respect to $x$.  Some care is
needed.  The
connected correlation functions that appear in (\ref{napp}) are
$O(\lambda)$ but depend on time, growing as the field grows away
from $\phi = 0$.  As is well understood, perturbation theory breaks down at
long
enough times.  However, for small times and weak coupling the first
term in (\ref{napp}) is reliable.  We see that, if
the Fourier transform ${\tilde W}(k;t)$ of $W(x;t)$ is dominated by
wave vectors $k(t) = O(\xi^{-1}(t))$ at time $t$, then
\begin{equation}
{\bar n}(t)\leq  O\bigg(\frac{1}{\xi (t)}\bigg).
\label{napp2}
\end{equation}
This shows that $\xi (t)$ sets the domain size.

A similar qualitative analysis can be attempted for the density
correlation functions, but with less success, in the absence of
further approximation.
To be concrete, let us consider the consequences of $p_{t}[\Phi]$ being {\it
Gaussian}, for
which the approximate equality in (\ref{n11}) becomes {\it exact}.
That this is not a frivolous exercise in solving
what we can solve but a representation of reasonable dynamics will be
shown, in part, later, for the 'slow-roll' dynamics that we
shall adopt.  For the moment we  take it for granted.

Specifically, suppose that (still for the one-dimensional case)
$\Phi$ is a Gaussian field for which
\begin{equation}
\langle\Phi (x)\rangle_{t} = 0 =
\langle\Phi (x)\Phi '(x)\rangle_{t},
\label{g11}
\end{equation}
and that
\begin{equation}
\langle\Phi (x)\Phi (y)\rangle_{t} = W (|x - y|;t).
\label{g22}
\end{equation}
All other connected correlation functions are taken to be zero.
Then all ensemble averages are given in terms of $W(r;t)$ which we
have have seen to be the equal-time Wightman function,
\begin{equation}
W (|x - y|;t) = \langle\phi (x, t)\phi (y, t)\rangle
\label{w1}
\end{equation}
with
the given initial conditions.  In our case, where we shall assume thermal
equilibrium initially, this is the usual thermal Wightman function.
It is then straightforward to see that, if
\begin{equation}
f(r;t) = \frac{W(r;t)}{W(0;t)}
\label{f}
\end{equation}
then
\begin{equation}
{\bar n}(t) =
\frac{1}{\pi}(-f''(0;t))^{\frac{1}{2}}.
\label{nii}
\end{equation}
in agreement with (\ref{napp}).  On using the same exponentiation that
$\langle\delta [\Phi (x)]\rangle_{t}$ equals
$\int d\!\!\!/\alpha\langle e^{i\alpha\Phi (x)}\rangle_{t}$
it takes only a little manipulation to cast
the correlation function $C(|x|;t)$ of (\ref{c1}) in the form
\begin{equation}
C(r;t) = \frac{\partial h(r; t)}{\partial r},
\label{C1}
\end{equation}
where
\begin{equation}
h(r;t) = \frac{-f'(r;t)}{2\pi\sqrt{1 - f^{2}(r;t)}}.
\label{h}
\end{equation}

If we are given $C(r;t)$, and wish to infer $f(r;t)$, and hence its
power spectrum, we integrate equation (\ref{h}) as
\begin{equation}
f(r;t) = \sin\bigg[\frac{\pi}{2} - 2\pi\int_{0}^{r}dr'\,h(r';t)\bigg]
\end{equation}
in which $h(r;t)$ is, in turn, derived from $C(r;t)$ through
equation (\ref{C1}).  For example, for the Poisson distribution of zeroes
given earlier in (\ref{poisson}), we find
\begin{equation}
f(r;t) = sin\bigg(\frac{\pi}{2}\,e^{-2r/\xi (t)}\bigg)
\end{equation}

However, the situation is usually the converse, with the dynamical model
predicting $f(r;t)$, from which $C(r;t)$ follows.  As a concrete example,
suppose that
\begin{equation}
W(r;t) = \int d\! \! \!/ k\,e^{ikx}\,{\tilde W}(k;t)
\end{equation}
where
\begin{equation}
{\tilde W}(k;t) = \delta (k^{2} -k_{0}^{2}(t)).
\end{equation}
That is, there is only one wavelength in the model.  Then
\begin{eqnarray}
f(r;t) &=&
\cos\,(k_{0}(t)r),
\nonumber
\\
{\bar n}(t) &=& k_{0}(t)/\pi .
\end{eqnarray}
The resulting
\begin{equation}
h(r;t) =\frac{k_{0}(t)}{2\pi}{\rm sign}(\sin\,(k_{0}(t)r)),
\label{hr'}
\end{equation}
is exactly that of (\ref{hr}), in which the correlation function
$C(r;t)$ is a sum of delta-functions, corresponding to a {\it regular} array of
zeroes.

Now suppose that, instead of a single wavelength, $W(r;t)$ has
contributions from frequencies peaked about $k_0$, as
\begin{equation}
{\tilde W}(k;t) = \frac{1}{2}\bigg(
\exp\{-\frac{1}{2}(k-k_{0}(t))^{2}/(\Delta k_{0}(t))^{2}\} +
\exp\{-\frac{1}{2}(k+k_{0}(t))^{2}/(\Delta k_{0}(t))^{2}\} \bigg)
\end{equation}
In the limit $\Delta k_{0}(t)\rightarrow 0$ we recover the regular
array but, for $\Delta k_{0}(t)\neq 0$,
\begin{equation}
W(r;t) = \int dk\,\cos kx\,\exp\{-\frac{1}{2}(k-k_{0}(t))^{2}/(\Delta
k_{0}(t))^{2}\}
\end{equation}
giving
\begin{equation}
f(r; t) = (\cos (k_{0}(t)r))\exp\{-\frac{1}{2}r^{2}(\Delta k_{0}(t))^{2}\}.
\label{f2dis}
\end{equation}
The corresponding $h(r;t)$ is no longer that of (\ref{hr}), but
\begin{equation}
h(r) = \frac{k_{0}\sin(k_{0}r) + r(\Delta k_{0})^{2}\cos (k_{0}r)}
{\sqrt{1-(\cos^{2} (k_{0}r))\exp\{-r^{2}(\Delta k_{0})^{2}\}}}
\exp\{-\frac{1}{2}r^{2}(\Delta k_{0})^{2}\}.
\end{equation}
and we have dropped the time-labelling.  A cursory inspection shows
that $h(r)$ is damped periodic, and
the broadening of its derivative $C(r;t)$ from its $\delta$-function behaviour
when $\Delta
k_{0}(t)\neq 0$ measures the variance $\Delta\xi (t)$ in the separation
$\xi (t)$ between adjacent zeroes.  We see
that, for small $\Delta k_{0}(t)/k_{0}(t)$,
\begin{equation}
\frac{\Delta\xi (t)}{\xi (t)}\propto \frac{\Delta k_{0}(t)}{k_{0}(t)}.
\label{rms0}
\end{equation}
with a coefficient of proportionality of $O(1)$\footnote{It follows that
the effect of the variance is to increase the density
of zeroes to
${\bar n}(t) =\sqrt{k_{0}^{2}(t) +(\Delta k_0 (t))^{2}}/\pi$.}.
We note that, even when the variance $\Delta\xi (t)/\xi (t)$ is
large, so that the zeroes appear much more random, the Gaussian
behaviour is very different from the linear exponential behaviour of
the Poisson distribution noted earlier.  There is always
correlation.

\subsection{Monopoles}

The extension to $O(2)$ line zeroes is messy, but leads to no surprises.
Specifically, suppose that
\begin{equation}
\langle\Phi_{a}({\bf x})\rangle_{t} = 0 =
\langle\Phi_{a}({\bf x})\partial_{j}\Phi_{b}({\bf x})\rangle_{t},
\label{g1}
\end{equation}
and, further, that
\begin{equation}
\langle\Phi_{a}({\bf x})\Phi_{b}({\bf x}')\rangle_{t} = W_{ab}(|{\bf x} -{\bf
x} '|;t)
= \delta_{ab} W(|{\bf x} -{\bf x} '|;t),
\label{g2}
\end{equation}
is diagonal.  As before, all other connected correlation functions are taken to
be zero.

The density calculation proceeds as before.
It follows\cite{halperin,maz} that
\begin{equation}
{\bar n}(t) =
\frac{1}{2\pi}(-f''(0;t)).
\label{ni}
\end{equation}
where derivatives are taken with respect to $r = |{\bf x} -{\bf x} '|$.
The zero density- density correlation
function
is more complicated than for the kinks, albeit still in terms of
$h(r;t)$ of (\ref{h}), as\cite{maz}
\begin{equation}
C(r;t) = \frac{2}{r}h(r;t)h'(r;t).
\label{C2}
\end{equation}

If $C(r;t)$ is given, we can integrate equation (\ref{C2}) to obtain
$f(r;t)$ as
\begin{equation}
f(r;t) = \sin\bigg[\frac{\pi}{2}
-
2\pi\int_{0}^{r}dr\,\bigg(-\int_{r}^{\infty}dr'\,r'\,C(r';t)\bigg)^{1/2}\bigg],
\label{fC}
\end{equation}
on incorporating the conservation of topological charge.

However, we are more interested in calculating $C(r;t)$ for specific $f(r;t)$.
As for kinks above, the simplest case corresponds to taking
\begin{equation}
W(r;t) = \int d\! \! \!/ ^{2}k\,e^{i{\bf k}.{\bf x}}\,{\tilde W}({\bf k};t)
\end{equation}
in which there is only one wavelength
\begin{equation}
{\tilde W}({\bf k};t) = \delta ({\bf k}^{2} -k_{0}^{2}(t)).
\end{equation}
where $k_{0}(t)$ depends on time $t$. As a result
\begin{equation}
f(r;t) = J_{0}(k_{0}(t)r),
\end{equation}
\begin{equation}
{\bar n}(t) = \frac{k_{0}^{2}(t)}{\pi}
\end{equation}
and, from (\ref{h}),
\begin{equation}
h(r;t) = \frac{J_{1}(k_{0}(t)r)}{2\pi\sqrt{1-J_{0}^{2}(k_{0}(t)r)}}.
\end{equation}
Because of its rotational invariance $C(r;t)$
 of (\ref{c2}) cannot
show the $\delta$-function behaviour of its counterpart on the line,
but regularity is implicit in the strong regular oscillatory
peaking of $C$ that comes from the Bessel functions in the numerator.

If, instead of a single wavelength, $W(r;t)$ has
contributions from frequencies peaked about $k_0$ with variance $\Delta
k_0$,
then the secondary peaking is reduced.  Specifically, consider
\begin{equation}
W(r;t) = \int dk\,J_{0}(kr) \,\exp\{-\frac{1}{2}(k-k_{0}(t))^{2}/(\Delta
k_{0}(t))^{2}\}.
\label{w2dis}
\end{equation}
In fact, unlike the one-dimensional case, for which $W(r;t)$ and
$f(r;t)$ of (\ref{f2dis}) retain their oscillatory behaviour, when $\Delta
k_{0}(t)/k_{0}(t)$
is large enough $W(r;t)$ does {\it not} oscillate or even change sign.
This is understood as indicating a less regular distribution of monopoles.
The dispersion in wavelength $\Delta k_{0}(t)$ leads to a variance $\Delta \xi$
in their separation that we expect to satisfy
\begin{equation}
\frac{\Delta\xi (t)}{\xi (t)}\propto \frac{\Delta k_{0}(t)}{k_{0}(t)}.
\label{rms2}
\end{equation}
In itself this short-range behaviour of $f(r;t)$ is enough to give the variance
in winding number
through a surface of perimeter $L$ as $(\Delta N)^{2} =
O(L)$ for the reasons given earlier.

\subsection{Vortices}

Finally, for line zeroes the Gaussian approximation has yet more complicated
consequences.
As for monopoles, we assume diagonal correlation functions
\begin{equation}
\langle\Phi_{a}({\bf x})\Phi_{b}({\bf x}')\rangle_{t} = W_{ab}(|{\bf x} -{\bf
x} '|;t)
= \delta_{ab} W(|{\bf x} -{\bf x} '|;t),
\label{gg3}
\end{equation}
from which $f(r;t)$ can be defined as in (\ref{f}), and vanishing
field expectation value and and the  independence of the field and
its derivatives
\begin{equation}
\langle\Phi_{a}({\bf x})\rangle_{t} = 0 =
\langle\Phi_{a}({\bf x})\partial_{j}\Phi_{b}({\bf x})\rangle_{t},
\label{gg4}
\end{equation}
As might have been anticipated, ${\bar n}(t)$ is as for monopoles
\begin{equation}
{\bar n}(t) =
\frac{1}{2\pi}(-f''(0;t)),
\label{ni2}
\end{equation}
whereas\cite{halperin,maz} the transverse and longitudinal parts of the density
correlation (\ref{ddc}) are, still in terms of $h(r;t)$,
\begin{equation}
A(r;t) = \frac{2}{r}h(r;t)h'(r;t)
\label{a}
\end{equation}
(the same in form as $C(r;t)$ of (\ref{c2})), and
\begin{equation}
B(r;t) = \frac{2}{r^{2}}h^{2}(r;t) > 0.
\label{b}
\end{equation}
We note that $B$ is {\it positive}.
The conservation law (\ref{Cint3b}) follows, as
\begin{equation}
\int d^{3}x\,(2A + B)\propto\int_{0}^{\infty}dr\,\frac{d}{dr}\bigg(rh^{2}\bigg)
=0
\end{equation}

Given either $A(r;t)$ or $B(r;t)$ we can reconstruct $f(r;t)$.  For
given $A(r;t)$, $f(r;t)$ has a similar form to that of (\ref{fC})
(except for the absence of charge conservation).  More simply, for
given $B(r;t)$,
\begin{equation}
f(r;t) = \sin\bigg[\frac{\pi}{2}
- \pi\sqrt{2}\int_{0}^{r}\,dr'\,r'(2B(r';t))^{1/2}\bigg],
\label{fB}
\end{equation}

If, as before, we asssume a single wavelength, then
\begin{eqnarray}
W(r;t) &=& \int d\! \! \!/ ^{3}k\,e^{i{\bf k}.{\bf x}}\,\delta ({\bf k}^{2}
-k_{0}^{2}(t))
\nonumber
\\
&=&\frac{\sin (k_{0}(t)r)}{k_{0}(t)r}\equiv {\rm sinc}\,( k_{0}(t)r)
\label{sing}
\end{eqnarray}
It follows that
\begin{equation}
{\bar n}(t) = \frac{k_{0}^{2}(t)}{6\pi}
\label{ent}
\end{equation}
and
\begin{equation}
h(r;t) = \frac{\sin (k_{0}(t)r) - k_{0}(t)r\, \cos (k_{0}(t)r)}
{2\pi r\sqrt{k^{2}_{0}(t)r^{2} - \sin^{2}(k_{0}(t)r)}}
\end{equation}
from which $A(r;t)$ and $B(r;t)$ can be constructed from above.
$A$ and $B$ again show periodic peaking, pointing to as regular a
distribution of line-zeroes as possible.
Specifically, when $k_{0}(t)r\gg 1$
\begin{equation}
A(r;t)\simeq \frac{-k_{0}(t)\sin 2k_{0}(t)r}{4\pi r^{3}}
\end{equation}
and
\begin{equation}
B(r;t)\simeq \frac{\cos^{2}k_{0}(t)r}{2\pi r^{4}}
\end{equation}
up to non-leading terms.
On the other hand, when $k_{0}(t)r\ll 1$
$A(r;t) = O({\bar n}^{2}(t))$
and negative, whereas
$B(r;t) = O({\bar n}^{2}(t)/k^{2}_{0}(t)r^{2})$.

Yet again, if, instead of a single wavelength, $W(r;t)$ has
contributions from frequencies peaked about $k_0$ with variance $\Delta
k_{0}(t)$ the secondary peaking is reduced.  If we take
\begin{equation}
W(r;t) = \int dk\,{\rm sinc}(kr) \,\exp\{-\frac{1}{2}(k-k_{0}(t))^{2}/(\Delta
k_{0}(t))^{2}\}.
\label{w3dis}
\end{equation}
then, as for monopole zeroes in two dimensions, once $\Delta
k_{0}(t)/k_{0}(t)$ is sufficiently large, $W(r;t)$ ceases to
oscillate or even to vanish.
Yet again we understand the variance in $k$ as inducing a variance $\Delta \xi$
in their separation approximately
satisfying
(\ref{rms2}).  However, the short-distance behaviour of $A$ and $B$
is unchanged qualitatively.

We shall see later, in our model
making, how distributions approximately of the
form (\ref{w3dis}) arise naturally.
However, there is another,
even simpler, and not wholly dissimilar way of introducing a dispersion about
the wavenumber
$k_0 (t)$ that has been adopted in a cosmological context\cite{andy2}.
Because of the similarities between it and the model that we shall
introduce later we shall look at it in some detail.
The work of \cite{andy2} essentially consists in taking
$W(r;t)$ as
\begin{equation}
W^{(n)}(r;t)\propto \int_{0}^{k_{0}(t)} dk\,{\rm sinc}(kr)
\,\bigg(\frac{k}{k_{0}(t)}\bigg)^{2+n}
\label{w3cut}
\end{equation}
describing fields with a variable power spectrum $k^n$, cut off at
$k =k_{0}(t)$.

For $n\geq 0$ the behaviour of the
normalised field correlation
functions $f^{(n)}(r;t) = W^{(n)}(r;t)/W^{(n)}(0;t)$ is determined
by the sharp cutoff, unlikely to be present in any realistic models.
However,  it
follows that the limit $n\rightarrow\infty$ reproduces the
single-mode result of (\ref{sing}).
The density ${\bar n}(t)$ is calculated easily as
\begin{equation}
{\bar n}(t) = \bigg(\frac{3+n}{5+n}\bigg)\frac{k_{0}^{2}(t)}{6\pi}
\label{ncut}
\end{equation}
reducing to (\ref{ent}) as $n\rightarrow\infty$, as it must.

As $n$ decreases the power in long wavelength modes increases.
For example, for $n = 0, -1, -2$
\begin{eqnarray}
f^{(0)}(r;t) &=& \frac{3}{(k_{0}(t)r)^{3}}(\sin(k_{0}(t)r) -
k_{0}(t)r\,\cos(k_{0}(t)r)
\label{fm1}
\nonumber
\\
f^{(-1)}(r;t) &=& \frac{2}{(k_{0}(t)r)^{2}}(1-\cos(k_{0}(t)r))
\\
f^{(-2)}(r;t) &=& \frac{1}{k_{0}(t)r}\,Si(k_{0}(t)r)
\nonumber
\label{fns}
\end{eqnarray}
The case $n=0$ corresponds to white noise on scales larger than
$k_{0}^{-1}$\footnote{White noise on all scales gives a $\delta$-function
for $W$.}.
We see immediately that $f^{(-1)}(r;t)$
of (\ref{fm1}) has double the period of the large-$n$ $W(r;t)$ of
(\ref{sing}), indicating a reduced density directly.

If we calculate the density-density anticorrelation function
$A(r;t)$ for these $f^{(n)}$ and others we find
that, for small $r$ (where the effects of the sharp cutoff are less
apparent) it can be written as
\begin{equation}
A(r;t)\approx -{\bar n}^{2}(t)(a_n - b_n(k_{0}(t)r)^{2})
\end{equation}
where $a_n$ and $b_n$ are positive and, by extracting a factor ${\bar
n}^{2}(t)$ we are working at constant density.  The strength $a_n$ of the
anticorrelation is approximately constant and $O(1)$ for $n > 0$,
whereas for $n < 0$ it diminishes  rapidly, vanishing when
$n\approx -2$.  For $n > -2$
the range of the anticorrelation is determined by
$b_{n}^{-1}$.  Over the range $-2 < n < \infty$, $b_n$ is
approximately linear, growing from approximately zero at $n = -2$
with slope $O(1)$.  Thus, as $n$ becomes increasingly negative (but
$n > -2$) the range over which the strings have influence on one
another increases, but with diminishing strength.  For $n = -2$,
$k_{0}(t)$ ceases to be a dominant wavenumber and there is no
quantity to identify with a domain size.
We would interpret this as implying the
greatest variance in string-zero separations.

For large $r$, we note that
\begin{equation}
f^{(n)}(r;t) = O\bigg(\frac{1}{(k_{0}(t)r)^{2}}\bigg)
\label{fnas}
\end{equation}
for integer $n\geq -1$, and $f^{(-2)}(r;t) = O(1/(k_{0}(t)r))$.
The vanishing of ${\bar n}(t)$ at $n = -3$ is a consequence of the
infrared divergence at this value, but can be ignored since it is
difficult to see how values of $n < -2$ could arise from realistic dynamics.

With these examples behind us we are now ready to attempt to make
predictions in a simple dynamical model of Gaussian fluctuations.
\section{Gaussian Dynamics from an Instantaneous Quench and its Coarsening}

In  practice, our ability to construct $p_{t}[\Phi ]$ or,
equivalently, the field correlation functions over the whole
timescale $t > t_0$ from initial quantum fluctuations to late time
classicality is severely limited.  In particular, a Gaussian
approximation can have only a limited applicability.  To see what
this is, it is convenient to divide time
into four intervals, to each of which we adopt a different approach.
With  $M$ setting the mass-scale, there is an initial period
$t_0 < t < t_i = O(M^{-1})$, before which the field
is able to respond to the quench, however rapidly  it is
implemented, and which we can largely ignore.
Assuming weak coupling, of which more later, this is followed by an
interval $t_i < t < t_{sp}$ in which, provided the quench is
sufficiently rapid, domains in field phase form and
grow.  For the reasons given above vortices (or monopoles) will appear and be
driven
apart by the coalescence of these domains.  For the quartic  winebottle
potential $V(\phi
)$ with minima at $|\phi |=\phi_0$ domain growth begins to stop
once the field has reached its spinodal value $|\phi|^2 =
\phi^{2}_{0}/3 = O(M^2 /\lambda
)$ at the ring of
inflection $V''(\phi ) = 0$.
This occurs at times peaked around some $t = t_{sp}$.
To estimate $t_{sp}$ we observe that, for short times and weak coupling,
the long wavelength modes with wavenumber $k\approx 0$  grow
as ${\tilde \phi}_{{\bf k}}(t) = O(Me^{Mt})$  as they fall from the top of the
potential hill.
Thus $t_{sp}$ is given in terms of the coupling strength by
\begin{equation}
\exp\{-2Mt_{sp}\} = O(\lambda)
\label{tsp}
\end{equation}
or equivalently, $t_{sp} = O(M^{-1}\ln (1/\lambda))$.  We assume that
$\lambda$ is sufficiently small that $t_{sp}$ is significantly
larger than $t_i $.  The defects are beginning to freeze in and
fluctuations are now too small to undo
them.
In the third period, beginning at $t_{sp}$, the
field magnitude relaxes dissipatively to the ground state values and the
vortices
complete their freezing in. Finally, in the
last period, the vortices behave semiclassically.

We shall see that our ability to calculate from first principles is limited
to the second period $t_i < t <
t_{sp}$.  On the (as yet unproven) assumption that the distribution of
relevant zeroes at time $t_{sp}$ is left largely unchanged by their
final freezing in,
this distribution of vortices can then be taken as initial data for
the final evolution of the network.  Similar considerations apply to monopoles.

It is not difficult to justify our adoption of Gaussian field
fluctuations for this second period $t_i < t <
t_{sp}$  of vortex production.
We have already assumed that the initial conditions correspond
to a disordered state.  In the absence of any compelling evidence to
the contrary we achieve this by adopting thermal equilibrium at
a temperature $T$ higher than the critical temperature $T_{c}$
for $t < t_{0}$, as we anticipated earlier.
That is, the action $S_0$ of (\ref{S-0}) that characterises the initial
conditions
is
\begin{equation}
S_0[\phi_3] = \int d^{D+1}x \biggl [
\frac{1}{2} (\partial_{\mu} \phi_{3 \, a} )(\partial^{\mu} \phi_{3 \,
a} ) - \frac{1}{2} m(T)^2 \phi_{3 \, a}^2 -\frac{1}{4}\lambda_{0}(\phi_{3 \,
a}^2)^{2}
\biggr ].
\label{S-00}
\end{equation}
with $m^{2}(T)>0$.
If $\lambda_{0}$ is weak the resulting field distribution is
approximately Gaussian, and it is little loss to take it to be
exactly Gaussian, $\lambda_{0}=0$.  As we shall see later,
initial conditions generally give slowly varying behaviour in the
correlation function, in contrast to the rapid variation due to the
subsequent dynamics, and calculations are insensitive to them.

In order to have as simple a change of phase as possible, we
assume an idealised quench,
in which, at
$t = t_{0}$, $m^{2}(t)$ changes sign everywhere.  Most simply, this change in
sign in
$m^{2}(t)$ can be interpreted as due to a reduction in temperature.
Even more, at first we further simplify our calculation by assuming that, for
$t > t_{0}$, $m^{2}(t)$
takes the {\it negative} value $m^{2}(t) = - M^2 <0$ {\it
immediately}, where $ - M^2$ is the mass parameter of the (cold)
relativistic Lagrangian.  That is,
the potential at the origin has been {\it instantaneously} inverted,
breaking the global $O(2)$ symmetry.   If, as we shall
further assume,  the $\lambda |\phi|^{4}$ field coupling
is very weak then,
for times $M(t -t_{0}) <
\ln(1/\lambda)$, the $\phi$-field, falling down the hill away from
the metastable vacuum, will not yet have experienced the upturn of
the potential, before the point of inflection and we can set
$\lambda =0$.
Thus, for these small time intervals,  $p_{t}[\Phi ]$ is Gaussian, as required.
Henceforth, we take $t_{0} = 0$.

For such a weakly coupled field
the onset of the phase transition at time $t=0$ is characterised by
the instabilities of long wavelength fluctuations permitting the growth of
correlations. Although the initial
value of $\langle \phi \rangle$ over any volume is zero, we
anticipate that
the resulting evolution will lead to
domains of constant $\phi $ phase, whose boundaries will
trap vortices.
This situation of inverted harmonic oscillators was studied many
years ago by Guth and Pi\cite{guth} and Weinberg and Wu\cite{weinberg}.
In the context of domain formation, we refer to
the recent work of Boyanovsky et al.\cite{boyanovsky}, and our own\cite{alray}.

Consider small amplitude fluctuations of $\phi_a$, at the top of the
parabolic potential hill.  Long wavelength fluctuations, for which $|{\bf
k}|^2 < M^2$, begin to grow exponentially. If their growth rate
$\Omega (k) = \sqrt{M^2 - |{\bf k}|^2}$ is much slower than
the rate of change of the environment,
then those long wavelength modes are unable to track the quench.
Unsurprisingly, the time-scale at which domains appear in this
instantaneous quench is $t_i = O(M^{-1})$. As long as the time taken
to implement the quench is comparable to $t_i$
and less than $t_{sp} = O(M^{-1}\ln (1/\lambda ))$
the approximation is relevant.  We shall relax the condition  in the next
section.

For the moment we ignore the effect of the interactions that
stabilise the potential and permit the zeroes corresponding to the
defects to freeze in.  In this free-roll period,
$W(r;t)$ has to be built from the modes ${\cal
U}^{\pm}_{a,k}$, ${\cal
U}^{+}_{a,k} =({\cal
U}^{-}_{a,k})^*$,  satisfying the equations of
motion
\begin{equation}
\Biggl [ \frac{d^2}{dt^2} + {\bf k}^2 + m^2(t) \Biggr ] {\cal
U}^{\pm}_{a,k} =0,
\label{mode0}
\end{equation}
For the idealised case proposed above $m^{2}(t)$ is of the form
\begin{eqnarray}
m^2(t) &=& m_0^2 >0 \, \,\, \mbox{if $t<0$,}
\nonumber
\\
&=& - M^2 <0  \, \, \mbox{if $ t>0$}.
\label{modes}
\end{eqnarray}
If we make a separation
into the unstable long wavelength modes, for which $|{\bf k}|<M$,
and the short wavelength modes $|{\bf k}|>M$, then  $W(r;t)$
is the real quantity
\begin{eqnarray}
W(r;t) &=& \int d \! \! \! / ^D k
\, e^{i {\bf k} . {\bf x} } C(k)
{\cal
U}^{+}_{a,k}(t){\cal
U}^{-}_{a,k}(t)
\label{Us}
\\
&=& \int_{|{\bf k}|<M} d \! \! \! / ^D k
\, e^{i {\bf k} . {\bf x} } C(k)
\biggl [ 1+ A(k)(\cosh(2\Omega (k)t) - 1 ) \biggr ]
\nonumber
\\
&+& \int_{|{\bf k}|>M} d \! \! \! / ^D k
\, e^{i {\bf k} . {\bf x} } C(k)
\biggl [ 1+ a(k)(\cos(2w(k)t) - 1 ) \biggr ]
\label{Ga}
\end{eqnarray}
with no summation over $a$, $r = |{\bf x}|$ and
\begin{eqnarray}
\Omega^2(k) &=& M^2 - |{\bf k}|^2,
\nonumber
\\
w^{2}(k) &=& -M^2 + |{\bf k}|^2,
\nonumber
\\
A(k) &=& \frac{1}{2} \biggl (
1+ \frac{\omega ^2(k) }{\Omega^2(k)} \biggr ),
\nonumber
\\
a(k) &=& \frac{1}{2} \biggl (
1- \frac{\omega ^2(k) }{w^2(k)} \biggr ).
\label{defs}
\end{eqnarray}
The initial conditions are encoded in ${\cal C}(k)$, which takes the familiar
form
\begin{equation}
{\cal C}(k) = \frac{1}{2\omega (k)}\coth(\beta_{0}\omega (k)/2)
\label{cr}
\end{equation}
in which $\omega^{2}(k) =  |{\bf k}|^2 + m_{0}^{2}$.

In the single-mode and other approximations adopted earlier the folly of
counting {\it all}
zeroes was
not apparent.  It is now, in the presence of the
ultraviolet divergence of $W(r;t)$ at $r=0$ in all dimensions.
None of the expressions given above is well-defined.  To identify
which zeroes will turn into our vortex network requires
coarse-graining.

The way to do this is determined by the dynamics.
Firstly, we note that if $\Phi$ is Gaussian,
then so is the  coarsegrained field on scale $L$,
\begin{equation}
\Phi_{L}({\bf x}) = \int d^{D}x'\,I(|{\bf x} - {\bf x}'|)\Phi ({\bf x}'),
\label{phil}
\end{equation}
where $I(r)$ is an indicator (window) function, normalised to unity,
which falls off rapidly for $r>L$.
The only change is that (\ref{g2}) is now replaced by
\begin{equation}
\langle\Phi_{L,a}({\bf x})\Phi_{L,b}({\bf x}')\rangle_{t} = W_{L,ab}(|{\bf x}
-{\bf x} '|;t)
= \delta_{ab} W_{L}(|{\bf x} -{\bf x} '|;t),
\label{gn2}
\end{equation}
where $W_{L}=\int\int IWI$ is now cut off at distance scale $L$.
$W_{L}(0;t)$, its derivatives, and all relevant quantities constructed
from $W_{L}$  are ultraviolet {\it finite}.  The distribution of
zeroes, or line zeroes, of $\Phi_L$ is given in terms of $W_{L}$ as
in the previous section.

Choosing $L = M^{-1}$ solves all our problems simultaneously.
At wavelengths $k^{-1} < L$ ({\it i.e.}$k > M$) the field fluctuations
are oscillatory, with time scales $O(M^{-1})$.  Only those
long wavelength fluctuations with $k^{-1} > L$ have the steady exponential
growth
that can lead to the field migrating on larger scales to its
groundstates.  Further, as the field settles to its groundstates,
the typical vortex thickness is $O(M^{-1})$, and we only wish to
attribute {\it one} zero to each vortex cross-section (or one zero
to each monopole).  By taking $L = M^{-1}$ we
are choosing not to count zeroes within a string, apart from the
central core.  We shall see later that it is not crucial to take $L =
M^{-1}$ exactly, but sufficient to take $L = O(M^{-1})$.

We are now in a position to evaluate $p_t[\Phi]$, or rather $W_{L}(r;t)$,
which we now write as $W_{M}(r;t)$ for $t > 0$, where
\begin{equation}
 W_{M}(r;t) = \int_{|{\bf k}|<M} d \! \! \! / ^D k
\, e^{i {\bf k} . {\bf x} }{\cal C}(k)
\biggl [ 1+ A(k)(\cosh(2\Omega (k)t) - 1 ) \biggr ]
\label{Gl}
\end{equation}
or,equivalently
\begin{equation}
 W_{M}(r;t) = \int_{|{\bf k}|<M} d \! \! \! / ^D k
\, e^{i {\bf k} . {\bf x} }{\cal C}(k)
\biggl [ 1+ \frac{1}{2}A(k)(e^{\Omega (k)t} -e^{-\Omega (k)t})^{2} \biggr ]
\label{Gla}
\end{equation}
and
calculate the density of zeroes accordingly.
Even though the approximation is only
valid for small times, there is a regime $Mt\ge 1$, for small couplings, in
which
$t$ is large enough for $ e^{Mt}\gg 1$ and yet $Mt$ is still smaller than
$Mt_{sp} = O(\ln (1/\lambda))$
when the fluctuations begin to reach the spinodal point on
their way to the ground-state manifold.
In this regime the exponential term in the integrand dominates and
\begin{eqnarray}
W_{M}(r;t)&\simeq& \frac{1}{2}\int_{|{\bf k}|<M} d \! \! \! / ^D k
\, {\cal C}(k)A(k)
e^{i {\bf k} . {\bf x} }
\;e^{2\Omega (k)t}
\\
&=& \frac{1}{2(2\pi)^{D}}\int_{|{\bf k}|<M}\,d\Omega\, dk\,k^{D-1}\;e^{2\Omega
(k)t}
\, {\cal C}(k)A(k)
e^{i {\bf k} . {\bf x} }
\end{eqnarray}

For the moment we assume that $M$ and $m_{0}$ are comparable. In our
context of initial thermal equilibrium at temperature $T_0$ this corresponds to
beginning the quench from well above the transition if, as earlier, we identify
$m_0$ with the thermal mass at temperature $T_0$, most simply
approximated to one loop by
\begin{equation}
m_0^2 = -M^2\bigg(1-\frac{T_{0}^2}{T_c^2}\bigg)
\label{mth}
\end{equation}
where $T_c = O(M/\sqrt{\lambda})$ in the same approximation.  The
definition (\ref{mth}) needs some qualifications, but for the
moment we accept it as it stands
.  We shall
consider the effect of modifying it later.
For $T_0 > T_c$, and weak coupling
${\cal C}(k)$ of (\ref{cr}) is approximately
\begin{equation}
{\cal C}(k) = \frac{T_0}{{\bf k}^2 +m_{0}^{2}}.
\label{cr1a}
\end{equation}

The integral at time t is then dominated by the
peak in $k^{D-1} e^{2\Omega (k)t}$ at $k$ around $k_0$, where
\begin{equation}
t k_0^2 = \frac{D-1}{2} M \biggl ( 1 + O \biggl ( \frac {1}{Mt}
\biggr ) \biggr ).
\label{k0}
\end{equation}
To check the consistency of the assumption that ${\cal C}(k)$ and $A(k)$ are
slowly varying in comparison
to this peak we observe that, at the largest times of interest to us,
$t=O(t_{sp})$
\begin{equation}
k_0^2 = O\bigg(\frac{M}{t_{sp}}\bigg) \ll M^2
\label{k00}
\end{equation}
and thereby, for a quench from well above the transition, equally less than
$m_0^2$.
${\cal C}(k)$ of (\ref{cr1a}) can be approximated by
\begin{equation}
{\cal C}(k) = \frac{T_0}{m_{0}^{2}}.
\label{cr2}
\end{equation}
The effect of changing the initial
thermal conditions
is only visible in the $O(1/Mt)$ term.
Since the overall normalisation of $W(r;t)$ is irrelevant, both ${\cal C}(k)$
and
$A(k)\approx 1$ can be factored out.  For weak coupling we recover what would
have been our first naive guess
for the coarse-grained correlation
function $\langle\phi_{L,a}({\bf r},t)\phi_{L,b}({\bf
r}',t)\rangle$
based on the growth of the unstable modes $\phi_{a}({\bf
k},t)\simeq e^{\Omega (k)t}$ alone,
\begin{equation}
W_{M}(r;t)\propto\int_{|{\bf k}|<M} d \! \! \! / ^D k
\, e^{i {\bf k} . {\bf x} }
\;e^{2\Omega (k)t},
\label{Wapp}
\end{equation}
provided we have not quenched from too close to the transition and
$t>M^{-1}$ (preferably $t\gg M^{-1}$).  It is in this sense that our
conclusions are insensitive to the initial conditions.
However, we might hope that this insensitivity extends to
non-thermal initial conditions as long as we are far from the
transition.

We note that if we had coarse-grained to {\it longer} wavelengths ${|{\bf
k}|<M_{0}<M}$ then, once $Mt > M^{2}/M_{0}^{2}$ the dominant peak
lies inside the integration region and the results are insensitive
to the value of the cutoff.  Provided $M/M_{0}\simeq 1$ this will be
the case because of (\ref{k00}).

On approximating the peak around $k_0$ by a Gaussian (which assumes
this insensitivity to coarse-graining on somewhat larger scales)  we see that,
for monopoles in two dimensions $W_{M}(r;t)$ has the form of
(\ref{w2dis}),
\begin{equation}
W_{M}(r;t)\propto\int dk\,J_{0}(kr)
\,\exp\{-\frac{1}{2}(k-k_{0}(t))^{2}/(\Delta k_{0}(t))^{2}\},
\label{w2}
\end{equation}
where $k_0(t)$ is given by (\ref{k0}), and
\begin{equation}
\frac{\Delta k_{0}(t)}{k_{0}(t)} = \frac{1}{\sqrt{2}}.
\label{va2}
\end{equation}

Further, for the more interesting case of vortices in $D=3$ dimensions,
$W_{M}(r;t)$ can
be approximated by (\ref{w3dis}),
\begin{equation}
W_{M}(r;t)\propto \int dk\,{\rm sinc}(kr)
\,\exp\{-\frac{1}{2}(k-k_{0}(t))^{2}/(\Delta k_{0}(t))^{2}\},
\label{w3}
\end{equation}
where
\begin{equation}
\frac{\Delta k_{0}(t)}{k_{0}(t)} = \frac{1}{2}
\label{va3}
\end{equation}
is also large.
We interpret (\ref{va2}) and (\ref{va3}) as implying large
variance in the  separation of the zeroes that are defining our defects.

Although the forms (\ref{w2}) and (\ref{w3}) represent the spread
in wavenumber well they do so at the expense of a correct
description of the long wavelength modes.
$W_{M}(r;t)$ of (\ref{Wapp}) can be compared directly to the
empirical form of (\ref{w3cut}) by writing it (up to the
usual arbitrary normalisation) as
\begin{equation}
W_{M}(r;t)\simeq\int_{|{\bf k}|<M} dk\,{\rm
sinc}(kr)\bigg(\frac{k}{k_{0}(t)}\bigg)^{2}
\;e^{2\Omega (k)t}.
\label{Wand}
\end{equation}
In the terminology of (\ref{w3cut})
this shows that, for long wavelengths, the power is $n=0$, determined
entirely by the radial $k^{D-1}$ behaviour in $D=3$ dimensions.  Insofar as
distributions of vortices are determined by the power in long
wavelengths the peak at $k = k_{0}(t)$ may, to some extent, be approximated by
the cutoff at $k_{0}(t)$ of (\ref{w3cut}).  We shall return to this later.

Rather than evaluate correlation functions from
(\ref{Wapp}) directly it is convenient to approximate it differently.
It is not difficult to see that, for the values of $\Delta k_{0}(t)/k_{0}(t)$
given above, $W_{M}(r;t)$ approximately falls monotonically in $r$ to zero from
above (any oscillatory behaviour is of very small amplitude).  It is
a good approximation
if, in (\ref{Wapp}) we expand $\Omega (k)t$ as
\begin{equation}
\Omega (k)t\approx Mt - \frac{k^{2}t}{2M}
\label{Omegap}
\end{equation}
from which
\begin{equation}
W_{M}(r;t)\simeq e^{2Mt}\int_{|{\bf k}|<M} d \! \! \! / ^D k
\, e^{i {\bf k} . {\bf x} }
\;e^{-k^{2}t/M}.
\label{Wappp}
\end{equation}
This approximation maintains the peak in the integrand at $k_{0}(t)$
of (\ref{k0}).  Taking it as
it stands gives
\begin{equation}
f(r;t) = \exp\{-r^{2}M/4t\},
\label{fg}
\end{equation}
{\it independent} of dimension, on dropping the upper integration
bound\footnote{This is not a question of dropping coarse-graining
but merely approximating the integral.}.
As such it correctly reproduces the small-$r$ behaviour of $f(r;t)$
(and $f^{0}(r;t)$ of (\ref{fm1})) and, although
the simple Gaussian behaviour breaks down for $r^{2}M/4t\gg 1$, the
rapid large-$r$
falloff is qualitatively correct.

We conclude with one more observation.  Even in the disordered state
for $t<0$ there were zeroes (or lines of zeroes) induced by thermal
fluctuations.
Effectively, the equal-time two-point correlation function for $t<0$ is
\begin{equation}
W_{M}(r;t)\simeq T_0\int_{|{\bf k}|<M} d \! \! \! / ^D k
\, \frac{e^{i {\bf k} . {\bf x} }}{{\bf k}^2 + m_{0}^2} = \int_{|{\bf k}|<M} d
\! \! \! / ^D k
\,e^{i {\bf k} . {\bf x} }{\cal C}(k).
\label{Eqcorr}
\end{equation}
Because of equilibrium there is no time-dependence
in $W_{M}(r;t)$ of (\ref{Eqcorr}).  There is an implicit cutoff in
(\ref{Eqcorr}) at the thermal wavelength $\beta_0 =T_{0}^{-1}$, but we are not
interested in zeroes on a smaller scale than the same field correlation
length $M^{-1}$ that determines defect size.  In fact, assuming $m_0\approx
M$, it makes little difference whether we coarsegrain to $|{\bf k}| <
M$ or to $|{\bf k}| < m_0.$

Whatever, the surface density
of zeroes is
\begin{equation}
{\bar n}(t) = O(m_{0}^{2}) =O(M^{2}) .
\end{equation}
However, we should not think of these zeroes as the precursors of
the defects that appear after the transition.  They are totally
transient.  We see this by the presence of oscillatory factors $e^{i\omega
(k)\,\Delta t}$ in the two-point correlation function at unequal times $t,
t+\Delta t$.  In fact, the term (\ref{Eqcorr}) is the first term in
(\ref{Gl}) and (\ref{Gla}), in fact the {\it only} term in either of these
expressions when $t=0$.  The reason for
rewriting (\ref{Gl}) as (\ref{Gla}) was to discriminate between this
term (the $1$ in the square bracket of (\ref{Gl}) ) and the term
$-A(k)\approx -1$, in the same bracket, whose origin is the interference
between
exponentially increasing and decreasing terms in the mode evlution
and has nothing to do with thermal fluctuations directly.
Since the normalisation of $W_{M}(r;t)$ has
no effect on the density and distributions of zeroes and line zeroes
these  thermal fluctuations are suppressed, relative to the long
wavelength peak that shows the growth of domains, by a factor
$O(e^{-2Mt})$.  Thus, although the thermal fluctuations remain as
strong in absolute terms, their contributions to the counting of
zeroes, and hence the creation of
defects, vanishes rapidly, which is why we did not include them
earlier.

Similarly, had we chosen to coarse-grain on a somewhat shorter scale ${|{\bf
k}|<M_{0}}$, where $M_{0} > M$, then $W(r;t)$ would have acquired
some (finite) oscillatory terms from (\ref{Ga}) that in turn would
have been suppressed, relative to the long
wavelength peak, by a factor
$O(e^{-2Mt})$.  For large enough times $t_{sp}$ they would play no
contribution, provided $M/M_{0}\simeq 1$.  Taken with our earlier
observations on coarse-graining at larger scales than $M^{-1}$ we
see that coarse-graining on a scale comparable to the thickness of
cold defects is all that is required, with fine-tuning being
unnecessary.

\section{Slowing the Quench}

Our adoption of an instantaneous quench, in which $m^{2}(t)$ changes as in
(\ref{modes})
is obviously unrealistic.  Any change in the environment requires some
time $\tau$ to implement.
In the context of the flat spacetime free-field  approximation that
we have adopted so far we should replace $m^{2}(t)$ of (\ref{modes}) by an
effective $(mass)^{2}$ that interpolates between $m_{0}^{2}$ and
its cold value $-M^{2}$ over $\tau$.  If $\tau$ were so small that
$M\tau <1$ we expect no significant change, since the field would
not have responded in the time available anyway.  However, once
\begin{equation}
1 < M\tau < Mt_{sp}
\end{equation}
there will be an effect.  As the wavenumber $k$ of the field modes
increases towards $M$ the time available for their growth is
progressively reduced.  However, the long wavelength modes have the
same time to grow as before.  As a result, the peaking of the
integral of $W(r;t)$ will occur at a smaller value of $k$ for the
same time $t$ after the quench is begun, leading to a lower density of defects
than would have
occurred otherwise.
Alternatively, the field is correlated over larger distances than
would have happened otherwise.

To see how this occurs quantitatively it is not necessary to go
beyond simple approximations, in the absence of any compelling
reason to make a specific choice for $m^{2}(t)$. [Such a reason
occurs if we have a definite spacetime metric driving the
transition, as happens in inflationary models\cite{guth,devega}].

Consider the behaviour
\begin{eqnarray}
m^{2}(t) &=& m^{2},\,\,\,\,\,t<0
\nonumber
\\
&=& -M^{2}\frac{t}{\tau},\,\,0\leq t <\tau
\nonumber
\\
&=& -M^{2},\,\,\,\tau \leq t
\label{newm}
\end{eqnarray}
in which the quench is begun at time $t = 0$ and completed at time $t = \tau$.
The  rate of the  quench, $\tau^{-1}$ is assumed large, $\tau <
t_{sp}$.  That is,
\begin{equation}
1\leq M\tau < \ln(1/\lambda),
\end{equation}
so that the quench is complete before the field has experienced the turnup of
the potential towards its  minima.
As $\tau\rightarrow 0$ we recover the instantaneous behaviour examined earlier.

If the behaviour presented in (\ref{newm}) seems a little artificial in how the
behaviour before $t=0$
and after $t=0$ are connected we note that, from our previous discussion, the
behaviour of $m^2 (t)$
for $t\leq 0$ is largely irrelevant, as long as $m^2$ is comparable to $M^2$.
Only the behaviour for $t >0$ will be important.

The calculation of $f(r;t)$ now requires a proper solution of the mode equation
(\ref{mode0})
for $m^{2}(t)$ of (\ref{newm}),
in terms of which
$W(r;t)$ is again given by (\ref{Us}).
This is not possible to derive analytically, but it is not difficult to make
analytic
approximations.
As before we coarse-grain $W(r;t)$ to eliminate oscillatory modes, bounding
$|{\bf k}|$ by $M$.
Whereas modes of all wavelengths $|{\bf k}| < M$ begin to grow exponentially at
the same time
$t = 0$
for an instantaneous quench ($\tau = 0$), for $\tau\neq 0$ modes of wavenumber
$k$ do not begin
to grow until time $t_{\tau}(k)$, at which
\begin{equation}
k^{2} = -m^{2}(t_{\tau}(k))
\end{equation}
For the choice of $m^2$ above in (\ref{newm})
\begin{equation}
t_{\tau}(k) = \frac{\tau k^2}{M^2}
\end{equation}
As we noted earlier, although the very long wavelength growth is activated at
$t\approx 0$,
the shorter the wavelength (but still $|{\bf k}| < M$) the shorter the time the
modes have
available before their switchoff at $t = t_{sp} = O(M^{-1}\ln (1/\lambda ))$.

A crude, but helpful approximation, for times $M\tau <Mt <
ln(1/\lambda)$ is to mimic this effect by modifying (\ref{Wapp}) as
\begin{equation}
W_{M,\tau}(r;t)\simeq\int_{|{\bf k}|<M} d \! \! \! / ^D k
\, e^{i {\bf k} . {\bf x} }
\;e^{2\Omega (k)(t-t_{\tau}(k))},
\label{Wap}
\end{equation}
In the further approximation of (\ref{Omegap}), in which we expand
in powers of $k$,
\begin{eqnarray}
W_{M,\tau}(r;t)&\simeq&\int_{|{\bf k}|<M} d \! \! \! / ^D k
\, e^{i {\bf k} . {\bf x} }
\;e^{(t-t_{\tau}(k))(2M -k^{2}/M)}
\nonumber
\\
&=& e^{2Mt}\int_{|{\bf k}|<M} d \! \! \! / ^D k
\, e^{i {\bf k} . {\bf x} }
\;e^{-k^{2}(t+2\tau )/M}\bigg[1 + O(\frac{\tau k^4}{M^3})\bigg].
\label{Wa}
\end{eqnarray}

On comparing $W_{M,\tau}(r;t)$ of (\ref{Wa}) to that of (\ref{Wappp})
it follows that, at the level of the Gaussian approximation, the effect of
slowing the quench is to replace
$f(r;t)$ of (\ref{fapp}) by
\begin{equation}
f_{\tau}(r;t)\approx\exp\{-r^{2}M/4t_\tau\}
\end{equation}
where
\begin{equation}
t_\tau =  t+2\tau + O\bigg(\frac{\tau}{M(t + 2\tau)}\bigg)
\label{ttau}
\end{equation}
That is, in the first instance the effect of slowing the quench is
no more than to reproduce the behaviour of an instantaneous quench, displaced
in time by an interval
$2\tau$.  A more careful calculation would give slightly
different answers, but the qualitative conclusion that slowing the
quench reduces the defect density and hence makes the system look as
if it had begun earler, is correct.  In the absence of any reason to make a
particular
choice of quench, the result (\ref{ttau}) is adequate for our purposes.

\section{Freezing in the Vortices}

We understand the dominance of wavevectors about $k_{0}^{2} = M/t_\tau$ in
the integrand of $W(r;t)$ as defining a length scale
\begin{equation}
\xi_{\tau}(t) = O(\sqrt{t_{\tau}/M}),
\label{xit}
\end{equation}
 once $Mt > 1$, over which the independently varying fields $\phi_a$
are correlated in magnitude.
To be specific, let us take
$\xi_{\tau}(t) = 2 \sqrt{t_{\tau}/M}$, whence
\begin{eqnarray}
f_{\tau}(r;t)&=& \exp\{-r^{2}M/4t_\tau\}
\nonumber
\\
&=& \exp\{-r^{2}/\xi_{\tau}^{2}(t)\}.
\label{fapp}
\end{eqnarray}
With this definition, the number density of line zeroes at early times
is calculable from (\ref{ni}) as
\footnote{The density of monopoles in D=2 dimensions is identical.}
\begin{equation}
{\bar n}(t) = \frac{1}{\pi}
\frac{1}{\xi_{\tau}(t)^2},
\label{nt}
\end{equation}
both for monopoles and vortices,
permitting us to interpret $\xi_\tau  (t)$ as a correlation length
or, equivalently, a domain size for
phases.
Although the zeroes and line zeroes have yet to freeze in as defects, this
density of one
potential defect per few correlation areas is commensurate with the Kibble
mechanism cited earlier.

The zeroes that we have been tracking so far cannot yet be
identified with the vortices (and monopoles) which provide the
semiclassical network on the completion of the transition because the
field has not achieved its groundstate values.  In fact, and of
greater importance at this stage, it is not even uniform in magnitude.  Insofar
as a classical picture
is valid, the thermal fluctuations have kicked the field off the top of the
upturned potential hill in different directions at slightly
different times.  The work of Guth and Pi\cite{guth} shows that the
variance in this time is $\Delta t = O(M^{-1})$.  Thus, even if the
potential were a pure upturned parabola and the fields were to  stop
growing instantaneously at the moment that they
reached the spinodal value $V''(\phi_{sp}) = 0$ (defined in terms of
the non-Gaussian physical potential) it would take a
further time $O(\Delta t)$ before the field caught up in all places.

An instantaneous halt to an increasing field growth is a huge
oversimplification.
More realistically, as the fields approach their spinodal values the effective
$(mass)^2$
driving the expansion of the unstable modes decreases and the
expansion of the domains slows.  The energy of the fields in the
long wavelength modes will cause an overshoot towards the potential
bottom that, in the absence of dissipation, will be followed by a
rebound, and perhaps a repeat (or many repeats) of the whole
cycle\cite{boyanovsky}.  The details will
depend on the dissipation that the long wavelength modes endure.
This dissipation is necessary to enable the fields to relax to their
ground-state values, and for realistic fields has several sources.
In particular, the act of coarse-graining induces
dissipation\cite{boyanovsky2}, as
would the presence of other fields\cite{boyanovsky3}
\footnote{A further source of dissipation is an expanding metric,
inevitable in cosmological models.  This is beyond us here.}.

In this excursionary paper we shall just consider the rudiments of
the effect of the field self-interaction on slowing down and stopping domain
growth in an approximation in which dissipation is assumed to set in
rapidly at the spinodal field values.
At the simplest level this is, in some respects, like that
of slowing
the quench that we discussed previously, except that it occurs at the
end of the period of interest, rather than the beginning. As there, the long
wavelength modes are the least affected, the shorter wavelength
modes (but still with $k < M$) having less time to grow.  For
this reason the density of zeroes, now identifiable as defects, is
reduced.

As we have said, a full analysis is beyond the scope of this paper but, as with
the
slow quench, it is possible to provide a primitive approximation
in which we can see this explicitly.  This is enough for our present
purposes, for which the power in the long wavelength modes will turn
out to be the most relevant property that we need, and we hope that the crudity
of our approximations leaves this untouched.

If we wish to retain the Gaussian approximation for the field
correlation functions the best we can do is a mean-field
approximation, or something similar\footnote{But not a large-$N$
$O(N)$ limit, since we only have simple defects in $D=N$ and $D=N+1$
dimensions.}.
In this approximation equations of motion are linearised by the substitution
\begin{equation}
(\phi_{1}^{2}+\phi_{2}^{2})\phi_{1}\approx
(3\langle\phi_{1}^{2}\rangle
+\langle\phi_{2}^{2}\rangle )\phi_{1}
\end{equation}
and similarly for $\phi_2$.
Because of the diagonal nature of $W_{ab}$ this means that
\begin{equation}
(\phi_{1}^{2}+\phi_{2}^{2})\phi_{a}\approx
4W(0;t)\phi_{a}
\end{equation}
$W(r;t)$ still has the
mode decomposition of (\ref{Us}), but the modes ${\cal
U}^{\pm}_{a,k}$ now satisfy the equation (ignoring subtractions)
\begin{equation}
\Biggl [ \frac{d^2}{dt^2} + {\bf k}^2 + m^2(t) +
4\lambda \int d \! \! \! / ^D p
\, C(p)
{\cal
U}^{+}_{a,p}(t){\cal
U}^{-}_{a,p}(t)
\Biggr ] {\cal
U}^{\pm}_{a,k} =0,
\label{mode1}
\end{equation}

A detailed discussion of (\ref{mode1}) will be given elsewhere, but
a rough estimate of the effects of the interactions can be obtained
by ignoring self-consistency and retaining only the unstable modes
in the integral. For simplicity we revert to an
instantaneous quench $m^2(t) = -M^{2}(t)$ for $t>0$.
On using the definition for $\lambda$
in (\ref{tsp}), we might approximate (\ref{mode1}) in turn
by
\begin{equation}
\Biggl [ \frac{d^2}{dt^2} + {\bf k}^2 - M^{2} +
M^{2}\,e^{2M(t-t_{sp})}
\Biggr ] {\cal
U}^{\pm}_{a,k} =0,
\label{mode2}
\end{equation}
where, further, we have ignored all but the dominant exponential
behaviour due to a free-field roll.  Note that the
$\lambda$-coupling strength  has been absorbed in the definition of
$t_{sp}$, most justifiable for small $\lambda$.  Equation
(\ref{mode2})  has the correct qualitative behaviour in that there are growing
modes only for $t<t_{sp}$, after which all modes are oscillatory.

In this approximation a mode ${\cal
U}^{\pm}_{a,k}$ can only grow until time $t_{f}(k)$, defined by
\begin{equation}
k^2 = M^{2}\bigg( 1-e^{2M(t_{f}(k)-t_{sp})}\bigg).
\label{tfk}
\end{equation}
As anticipated, shorter wavelengths have less time to grow.
Linearising (\ref{tfk}) in the vicinity of $t_{sp}$ as
\begin{equation}
k^2 = 2M^{3}(t_{f}(k)-t_{sp})
\label{tfk2}
\end{equation}
shows that, beginning from an instantaneous quench, the mode with
wavenumber $k$ stops growing at time
\begin{equation}
\Delta t(k)\approx \frac{k^2}{2M^3}
\label{Dt}
\end{equation}
before the modes of longest wavelength stop.

Instead of the instantaneous quench, now let us
reintroduce a slower quench that takes time $\tau$ to
implement, along the lines of (\ref{newm}) and the approximations
that followed from it.  At the simplest level, the effect of the
back-reaction on the field correlation function $W_{M,\tau}(r;t)$ of
(\ref{Wap}) is
to replace it by
\begin{equation}
W_{M,\tau}(r;t)\simeq\int_{|{\bf k}|<M} d \! \! \! / ^D k
\, e^{i {\bf k} . {\bf x} }
\;e^{2\Omega (k)(t-t^{*}(k))},
\label{Wapb}
\end{equation}
where
\begin{eqnarray}
t^{*}(k)&\approx& t_{\tau}(k) +\Delta t(k)
\nonumber
\\
&=& \frac{\tau k^2}{M^2} +\frac{k^2}{2M^3}.
\end{eqnarray}
In the further approximation of (\ref{Omegap}), in which we expand
in powers of $k$,
\begin{equation}
W_{M,\tau}(r;t)\simeq\int_{|{\bf k}|<M} d \! \! \! / ^D k
\, e^{i {\bf k} . {\bf x} }
\;e^{(t-t^{*}(k))(2M -k^{2}/M)}.
\label{Wab}
\end{equation}
{}From this we derive an $f(r;t)$ of the form
\begin{equation}
f(r;t)\approx \exp\{-r^{2}M/4t_{\tau}\}
\label{ff}
\end{equation}
where $t_{\tau}$ is now given by
\begin{equation}
t_{\tau}\approx t + 2\tau + \frac{1}{M}.
\label{teef}
\end{equation}
rather than by (\ref{ttau}).
We stress that, even at this level of approximation (in which any
term can be multiplied by a coefficient $O(1)$), we have assumed
sufficiently weak coupling and a rapid enough quench that
\begin{equation}
\tau + \frac{1}{2M}<t_{sp}
\end{equation}
so that even the shortest wavelength considered has some time to grow.
The details of (\ref{ff}) and (\ref{teef}) should not be taken too
seriously, but the qualitative result that the effect of slowing the
quench is to give a time-delay of $O(\tau)$, and the effect of
back-reaction is to give a time-delay of $O(M^{-1})$ is as we would
expect.    A
more realistic calculation\cite{boyanovsky} based on (\ref{mode1})
would show that the field overshoots its
spinodal values, retreats and overshoots again in damped
oscillations.  However, the small-$r$ behaviour of (\ref{ff}) is
good enough for our immediate purposes. We shall return to the full
equation (\ref{mode1}) later.

\section{Densities and Correlations from a Quench}

In the approximation given above the effect of slowing the quench
(and taking the back-reaction into account) is just
to increase the correlation length and thereby reduce the
density of zeroes to one appropriate to an instantaneous transition that
happened
earlier and in which the field spontaneously stopped growing.
The only way the zero-density can
decrease without the background space-time expanding is by
zero-antizero annihilation.  For monopoles this
is simple removal of zeroes by superposition. For vortices it will
include the collapse of small loops of
zeroes.  The end result is to  preserve roughly
one string zero per coherence area,  a long held belief for whatever
mechanism.

Let us perform all our calculations at $t=t_{sp}$, for which
$\xi_{\tau}(t_{sp}) = \xi_{sp}$.
For small enough $\lambda$, the density at $t=t_{sp}$ is, with our
previous caveats about coefficients,
\begin{equation}
{\bar n}(t_{sp}) = \frac{1}{\pi}
\frac{1}{\xi_{sp}^2}\approx \frac{1}{4\pi}\frac{M^2}{(Mt_{sp} +
2M\tau +1)}\ll M^{2},
\label{nt2}
\end{equation}
We have retained the last term in the denominator as a reminder that
the approximation may work even when $Mt_{sp}>1$, but not too large.
This shows that the line zeroes (or
monopole zeroes) only create a small fraction of
space as false vacuum.  Equivalently, the typical separation of line zeroes (or
monopole zeroes) is significantly larger than the thickness of a
cold vortex (or monopole) once they have  frozen in.
 This is one reason why we can begin to consider these line zeroes (and point
zeroes) as serious candidates for vortices (and monopoles) even though the
field has not
fully relaxed to its ground states.

A further reason is that, even though the
spinodal values of field are at a fraction of $1/\sqrt{3}$ of the
distance to the bottom of the potential, this distance is very much
larger than the initial field fluctuations.
In consequence, the fluctuations in the field are now too
small to create new lines and points of false vacuum of any
substance and we can
safely say
that we have defects.
To see this, we observe that, initially, the probability
$p_{t}[\Phi]$, now independent of $t$, is expressable as
\begin{equation}
p[\Phi] = N\, e^{-\beta_0 H[\Phi]}
\end{equation}
$N$ is a normalisation factor and $H$ permits the expansion in
$\beta_{0} m_{0}$ \cite{ray} in terms of $S_{0}$ of (\ref{S-00})
\begin{equation}
H[\Phi] = S_0[\Phi] - \frac{1}{24}\beta_{0}^2\int d^{D}x\,\biggl (\frac{\delta
S_0}
{\delta\Phi}\biggr )^2
 + O(\beta_{0}^4 m_{0}^{4}).
\end{equation}
With calculational simplicity in mind we restrict ourselves to high
initial temperatures ($\beta_{0}m_{0}\ll 1$) for which it is
sufficient to retain only the first term, and take $\lambda_0 = 0$.

As with zeroes, fluctuations are defined with respect to some length
scale L, say.
The fields $\Phi_{L,a}$ coarse-grained to this length are defined in
(\ref{phil}).
The probability $p_{L}({\bar\Phi})$ that $\Phi_{L} = {\bar\Phi}$ can
be written as
\begin{equation}
p_{L}({\bar\Phi}) = N\, \exp(-{\bar\Phi}^{2}/2(\Delta_{L}\Phi)^{2})
\end{equation}
where $(\Delta_{L}\Phi)^{2})$ is the coarse-grained two-point
function
\begin{eqnarray}
(\Delta_{L}\Phi)^{2} &=& W_{L}(r;0)\simeq T_0\int_{|{\bf k}|<L^{-1}} d \! \! \!
/ ^D k
\, \frac{1}{{\bf k}^2 + m_{0}^2}.
\nonumber
\\
&=& \int_{|{\bf k}|<L^{-1}} d \! \! \! / ^D k\,{\cal C}(k)
\label{Eqcorr2}
\end{eqnarray}
for ${\cal C}(k)$ of (\ref{cr}), compatible with (\ref{Gl}).
For $m_{0}\approx M$ it follows\cite{ray} that, for $L = O(M^{-1})$
in $D=3$ dimensions
\begin{equation}
(\Delta_{L}\Phi)^{2}\simeq AMT_{0}
\end{equation}
where $A\simeq 10^{-1}$.

The condition that  there be no overhang is thus
\begin{equation}
(\Delta_{L}\Phi)^{2}\ll \phi_{0}^{2}
\label{cond}
\end{equation}
There is no difficulty in satisfying (\ref{cond}) for small
coupling, for which it just becomes $\lambda\ll 1$.

Although the strings of zeroes (or monopoles) are far apart they
still do not yet
quite provide the
semiclassical network of defects that can be used as an input for numerical
simulations since the fields have to relax dissipatively from their spinodal
values to  the true minima of the potential along the lines
suggested earlier.
Because they are too costly
in energy to be produced by fluctuations,  there will be some defect
annihilation, but no creation, in this final phase of freezing in
the defects and the ${\bar n}(t_{sp} )$ calculated previously is an
overestimate
of the string density at the end of the transition.  However, if it
were the case for vortices that all the string was in loops it is
difficult to see how infinite string could be created if all that
happens is that string is removed from the system, although it
cannot be precluded.
Since that is the main question that we shall be addressing here we
adopt the simplest assumption that, even if the
density ${\bar n}(t_{sp} )$ is an overestimate the distribution of
strings (i.e. the fraction of strings in loops, the index for length
distributions) is approximately unchanged by the final freezing in.  That is,
the
one-scale scaling regime remains valid till freeze-in.

 In that case the distribution of strings will then be as
above for $\xi = \xi (t_{sp})$, while
$\Delta\xi /\xi$ of (\ref{rms}) remains (albeit approximately) the
variation in domain size over which field {\it phase} is correlated.
Direct attempts to determine distributions from density correlation
functions are difficult.
For example,
the simple analytic form of (\ref{fapp}) enables us to calculate
the density correlation function $C(r;t_{sp})$ for monopoles in the
plane, and $A(r;t_{sp})$ and
$B(r;t_{sp})$ for vortices in three dimensions, up to exponentially small terms
in $Mt_{\tau}(t_{sp})$, as
\begin{equation}
C(r) =
A(r) = \frac{2}{\pi^{2}\xi^{4}_{sp}}
\frac{e^{-2r^{2}/\xi^{2}_{sp}}}{(1 - e^{-2r^{2}/\xi^{2}_{sp}})^{2}}
\biggl [(1 - e^{-2r^{2}/\xi^{2}_{sp}}) - 2\frac{r^{2}}{\xi^{2}_{sp}}
\biggr ] < 0,
\end{equation}
and
\begin{equation}
B(r) = \frac{2}{\pi^{2}\xi^{4}_{sp}}
\frac{e^{-2r^{2}/\xi^{2}_{sp}}}{(1 - e^{-2r^{2}/\xi^{2}_{sp}})} > 0,
\label{corr1}
\end{equation}
and we have dropped unneccessary time labels.

In units of ${\bar n}^{2}(t_{sp})$, the
anticorrelation is large.
In particular, a calculation of $C(r)$ or $A(r)$ for defect
separation $r$ gives
\begin{equation}
A(r)
= {\bar n}^{2}(t_{sp})\bigg[ -1 + O\bigg(\frac{r^{2}}{\xi^{2}_{sp}}\bigg)\bigg]
\label{ca}
\end{equation}
when $r < \xi_{sp}$, as happens for the single mode case.
Again, as in the single-mode case,
\begin{equation}
B(r) ={\bar n}^{2}(t_{sp})\biggl[\biggl(\frac{\xi^{2}_{sp}}{r^{2}}\biggr)
+ O(1)\biggr] > 0.
\end{equation}

Although (\ref{corr1}) is not wholly reliable for $r\gg\xi$, the
suggestion that $C$ or $A$ and $B$ fall off
very fast
is qualitatively correct, showing that $\xi_{sp}$ indeed sets the scale at
which strings see one
another. The best we can say is that, taken together these suggest a
significant
amount of string in small loops.

We have more success in using the correlation functions to determine
the variance in vortex winding number
$N_S$ through an open surface $S$ in the 1-2 plane, which we take
to be a disc of radius $R$.  There is a complication in that, as can
be seen from (\ref{corrr}),
$\rho_{3}$ counts zeroes weighted by the cosine of the angle with
which they pierce $S$, whereas winding number counts zeroes in $S$
without weighting.  The problem is therefore essentially a
two-dimensional problem.  Using the results of (\ref{per}) onwards, it is not
difficult to see
that, given the short range of the correlation functions, for the
density of zeroes
\begin{equation}
(\Delta N)_{S}^{2} = O({\bar n}^{2}\xi_{sp}^{3} R)= O(R/\xi_{sp}).
\end{equation}
Since $(\Delta N)_S$ is $(\Delta\alpha)_{S}/2\pi$, where
$(\Delta\alpha)_{S}$ is the variance in the field phase around the
perimeter $\partial S$ of $S$, this means in turn that
\begin{equation}
(\Delta\alpha)_{S}^{2} = O(R/\xi (t_{sp})).
\label{dal}
\end{equation}
If this were equally true for nonrelativistic vortex systems, then
$(\Delta\alpha)_{S}^{2}$ measures the variance in the supercurrent
produced by the quench\cite{zurek1} and such behaviour is
measurable, in principle.  In practice we do not yet know how to
construct the coefficient of $R$ for  non-relativistic fields, but
see Refs.\cite{rayp,timc}.

Returning to the main problem of the length distributions of vortices in three
dimensions,
it has not yet proved possible to turn expressions for the
$C_{ij}$ (i.e.$A$ and $B$) directly into statements about self-avoidance,
fractal dimension, or
whatever is required to understand the
resulting string network.
Fortunately, as a temporary expedient, we can work indirectly by
adapting the numerical results of \cite{andy2}, based on Gaussian
fields with the two-point function $W_{n}(r;t)$ of (\ref{w3cut})
that we considered earlier.
For $n=0$ this has the same
long-wavelength behaviour as our dynamical $W(r;t)$ of (\ref{Wand}) and
(\ref{Wappp}).
The results of \cite{andy2} for $n=0$ reproduce those of \cite{tanmay} on a
cubic lattice,
with its preponderence of open string.  This leads us to believe
that the simple dynamical model that we have proposed above
(given its assumptions for the freezing in of domains)
would
also lead to a large fraction of open string.

Extending the simulation to general $n$ for $W_{n}(r;t)$ of
(\ref{w3cut}) shows\cite{andy2} that, in
order to decrease the amount of string in open string, it is
necessary to {\it decrease} $n$.  However, although the
fraction of infinite string does diminish as $n$ becomes negative,
infinite string only seems to vanish in the pathological limit  $n\rightarrow
-3$,
at which there is an infrared divergence.  We see from
(\ref{ncut}) that the density of string vanishes in the same limit.
As long as there is a nonzero density of string, then some of
it is infinite.

The only way to increase the power in long wavelengths in our
dynamical model is to change ${\cal
C}(k)$ by imposing different initial conditions.
Altering ${\cal C}(k)$ from ${\cal C}(k) = O(k^0 )$ to ${\cal C}(k) = O(k^n )$
leads to a change in the power of the long wavelength modes from $n=0$ to $n$,
corresponding to a modification of
$W_{M}(r;t)$ of (\ref{Wand}) to
\begin{equation}
W_{M}(r;t)\propto\int_{|{\bf k}|<M} dk\,sinc(kr)\,k^{2+n}
\;e^{2\Omega (k)t}.
\label{Wand2}
\end{equation}

The effect of changing $n$ can be seen qualititively by expanding
$\exp\{\Omega
(k)t\}$ as in (\ref{Omegap}). $W_{M}(r;t)$ of (\ref{Wand2}) can then be
evaluated
explicitly, giving $f^{(n)}(r;t)$ as the confluent hypergeometric function
\begin{eqnarray}
f^{(n)}(r;t)&\approx&
\exp\{-r^{2}M/4t\}\,_{1}F_{1}\bigg(\frac{-n}{2};\frac{3}{2};\frac{r^{2}M}{4t}
\bigg)
\nonumber
\\
&=& _{1}F_{1}\bigg(\frac{n+3}{2};\frac{3}{2};-\frac{r^{2}M}{4t}\bigg)
\label{1F1}
\end{eqnarray}
on dropping the upper bound in the integral.

The densities obtained from the more complicated (\ref{1F1}) are
{\it exactly} those of
the more simple $W_{n}(r;t)$ given in (\ref{ncut}), if we identify $k_{0}^{2}$
as $M/2t$.
Further, for
$n>-2$ the integrand of $W(r;t)$ in (\ref{Wand2}) shows a peak whose
width increases as
\begin{equation}
\Delta k \propto\frac{1}{\sqrt{n+2}},
\end{equation}
implying a variation in domain size
that also increases as $n\rightarrow -2$, just as for the simpler case.

Since $_{1}F_{1}(0;3/2;r^{2}M/4t) =1$ identically, $n=0$ is a special case in
which we reproduce the Gaussian of (\ref{fg}).
Although this seems at variance with the behaviour given in (\ref{fnas})
for the simply cut-off power distribution, the exponential
behaviour was never supposed to be valid for very large $r$.
Despite that, for $n<0$ the
similarity is good, with the {\it same} power-law falloff for $n=-1$
and $n=-2$,
leading to anticorrelations over increasingly longer ranges.

Thus, if we can force ${\cal C}(k)$ to have negative $n$,
the numerical simulations of \cite{andy2} suggest that there will be
more string in loops, and less in infinite
string.
The reason why we have $n=0$ white noise in (\ref{Wand}) is that, when
quenching from
{\it well above} the transition, ${\cal C}(k) \approx
T_{0}/m_{0}^{2}$ is constant for the dominant integration region.
In attempting to alter ${\cal C}(k)$ we expect that,
if we were to begin closer to the transition
where the initial fluctuations are larger, we can make
$n$ negative.
On beginning a quench from thermal equilibrium, the most extreme case is
one  in which we start from a temperature so close to the transition that the
effective mass $m_0$ of (\ref{modes}) for $t < 0 $ is approximately
zero.
Let us first suppose that we could set it {\it identically} zero.  Then
\begin{equation}
{\cal C}(k) = \frac{T_{0}}{k^{2}}.
\label{cr3}
\end{equation}
rather than (\ref{cr2}).
The $k^{-2}$ behaviour cancels the $k^2$
behaviour coming from the  radial momentum integration and the power
in long wavelengths is increased from
$n = 0$ to $n= -2$ in (\ref{Wand2}).  We believe that this is the most
negative that $n$ can become.  Even if we had performed a more
honest calculation of (\ref{mode1}) there is no way for the
interactions to introduce the singular infrared behaviour of $n =
-3$ that is
necessary to prohibit the production of strings.
The simulations of \cite{andy2} suggest an approximate halving of
the fraction of open string for $n= -2$ from its original 80\% (on a
cubic lattice).
These numbers are not to be taken too seriously since, as we
commented earlier, they depend on the type of lattice\cite{mark}
(although the presence of open string does not).

As we had noted above, for such a ${\cal C}(k)$ there is no peaking in the
integrand around any wavenumber, and so no length that characterises
a domain size, although the string density is nonzero.
{}From (\ref{Eqcorr2}) we see that the same ${\cal C}(k)$ of
(\ref{cr3}) also
determines the initial field fluctuations.  However, the effect
there is small, contrary to naive expectation, provided we
coarse-grain to the scale appropriate to defects, $k < M$.  This is
the relevant scale, rather than the thermal wavelength $\beta_{0}$.
On coarse-graining  to $k<M$ the field
fluctuations, given by  (\ref{Eqcorr2}), still satisfy
$(\Delta_{L}\Phi)^{2}\simeq AMT_{0}$,
and the initial fluctuations will not populate the ground states for
weak coupling.

Our ability to make $n$ negative survives more realistic initial states.
The situation is more complicated in that the Gaussian
approximation based on (\ref{S-00}) breaks down\cite{ray} at temperatures
$T$ closer to $T_c$ than
\begin{equation}
\bigg|1-\frac{T^2}{T_c^2}\bigg| = O(\lambda ).
\end{equation}
Thus the minimum value of $m^{2}_{0}$ at which our approximations
have any hope is $m^{2}_{0} = O(\lambda M^2)$.  However, this is
small enough for (\ref{cr3}) to be a good approximation, provided
\begin{equation}
\lambda \,\ln(1/\lambda )\ll 1,
\end{equation}
as follows on implementing the approximation (\ref{Omegap}).
In practice quenching from so close is artificial,
and we should consider $n= -2$ as an unattainable lower bound.

However, this is not the end of the story.
The regular lattices of \cite{tanmay} and \cite{andy2} are artefacts
of the calculational scheme and may, of themselves, predict more
infinite string than is present if they try to emulate a continuous transition.
As we saw earlier, we do not have a regular
domain structure in our model but have domains with a large
variance
\begin{equation}
\frac{\Delta\xi}{\xi}\approx \frac{\Delta k}{k_{c}} = O(1)
\label{rms}
\end{equation}
independent of time.  Whatever the details, the dispersion
in $\xi$ is large.

Although a regular lattice does not imply exactly regular domains
(which we understand as characterising the spaces between the
defects) for high densities of the kind we have here they do imply a
greater regularity than we have.
In
numerical simulations of string networks
the inclusion of variance in the 'lattice' cell size
 shows\cite{andy} that, the greater the
variance, the more string is in small loops.  This can be understood
in the following way.  The strings generated by phase separation on
regular lattices are known to behave like random walks in $D=3$
dimensions to a very good approximation\cite{scherrer}. On a
rectangular lattice the fraction of string in loops is only about
20\%\cite{tanmay}. Increasing
the variance of the domain size increases the 'target area' that a
string must hit for a section of string to be deemed closed.  The
probability of finding loops therefore increases. The variance of
(\ref{rms}), if it could be carried over to \cite{andy} as it
stands,
 suggests much more string ({\it e.g.} twice or more) in small loops.
However, it is not easy to marry the somewhat different
distributions of this simulation to ours since ours is one of
overlapping domains (because of the continuous transition) rather
than a domain 'bubble' picture more appropriate to first-order
transitions that is more easily accomodated by \cite{andy}.

The end result is that there always seems to be infinite string from a
continuous
transition, even if less than we thought, although its confirmation will
require numerical
analysis along the lines of \cite{andy2}.  This will be performed
elsewhere when we have a better understanding of how back-reaction
stops domain growth\cite{boyanovsky} than the simple approximation presented
here.

\section{\bf Conclusions}

In this paper we have shown how global O(2) vortices (and monopoles) appear in
a
relativistic theory, at a
quench from the ordered to disordered state, as a consequence of the
growth of unstable Gaussian long wavelength fluctuations.

Assuming a simple freezing of defects
the resulting string (and monopole)
configurations scale as a function of the correlation length
$\xi
= O(\sqrt{t_{\tau}})$, for some effective time $t_{\tau}$, at a
fraction of a defect/correlation area.
The label $\tau$ characterises the time it takes the defects to
freeze in.  The slower the
quench (for weak coupling), the larger $t_{\tau}$, and hence the
lower the defect density.  For quenches from well above the
transition the results are insensitive to initial conditions, which
have been
assumed here to be thermal equilibrium, but possibly the result is more
general.
This is compatible with the Kibble mechanism for vortex production
due to domain formation upon phase separation by white noise
fluctuations (and similarly for
monopole production).
For a weak coupling theory the domain cross-sections are
significantly larger than a vortex (or monopole) cross-section at the largest
times for which the approximations are valid.  Moreover,  there is a large
variance in their size $\xi$, with $\Delta\xi /\xi = O(1)$.
Numerical simulations that relate loop distribution to the
variance in domain size of the long wavelength modes
suggest that  there is likely to be more string in
small loops than we might have anticipated on the basis of
simulations on regular lattices, although the match with
the simulations is not exact.

As we quench from closer to the transition and effectively reduce
$n$ we expect that the fraction of string in
loops increases (as the density of string decreases), but even then
it is never possible to enhance long wavelength
fluctuations to an extent that there is no infinite string after a
continuous transition.

We stress the crudity of some of the approximations that we have
made.  There is little difficulty, in principle and in practice, in
doing somewhat better.  However, without any specific choice of
initial conditions and mass evolution $m(t)$ to guide us, and
without better-matching numerical simulations, generic
results at this level are sufficient.  Improvements, motivated by
particular cosmological models, are being considered.

\section{Acknowledgements}

We would like to thank Tim Evans, Carl Bender, James Robinson and Andy
Yates at
Imperial College, Dan
Boyanovsky and Rich Holman at Pittsburgh and Woytiech Zurek at Los
Alamos for fruitful discussions, from which this work has matured.
G.K. would like to thank the Greek State Scholarship Foundation (I.K.Y.) for
financial support.

\end{document}